\documentclass[12pt, preprint]{emulateapj}
\usepackage{amsmath}
\usepackage{graphicx}

\shortauthors{L. BOCO ET AL.}
\shorttitle{TOPSEM}
\slugcomment{Accepted by ApJ}

\begin{document}

\title{TOPSEM, TwO Parameters Semi Empirical Model: galaxy evolution and bulge/disk dicothomy from two-stage halo accretion}

\author{L. Boco\altaffilmark{1,2,3}, A. Lapi\altaffilmark{1,2,3,4}, F. Shankar\altaffilmark{5}, H. Fu\altaffilmark{5}, F. Gabrielli\altaffilmark{1}, A. Sicilia\altaffilmark{1}}
\altaffiltext{1}{SISSA, Via Bonomea 265, 34136 Trieste, Italy}\altaffiltext{2}{IFPU-Institute for Fundamental Physics of the Universe, Via Beirut 2, 34014 Trieste, Italy}\altaffiltext{3}{INFN-Sezione di Trieste, via Valerio 2, 34127 Trieste,  Italy}\altaffiltext{4}{INAF/IRA, Istituto di Radioastronomia, Via Piero Gobetti 101, 40129 Bologna, Italy}\altaffiltext{5}{School of Physics and Astronomy, University of Southampton, Highfield, SO17 1BJ, UK}

\begin{abstract}
In recent years, increasing attention has been devoted to semi empirical, data-driven models to tackle some aspects of the complex and still largely debated topic of galaxy formation and evolution. We here present a new semi empirical model whose marking feature is simplicity: it relies on solely two assumptions, one initial condition and two free parameters. Galaxies are connected to evolving dark matter haloes through abundance matching between specific halo accretion rate (sHAR) and specific star formation rate (sSFR). Quenching is treated separately, in a fully empirical way, to marginalize over quiescent galaxies and test our assumption on the sSFR evolution without contaminations from passive objects. Our flexible and transparent model is able to reproduce the observed stellar mass functions up to $z\sim 5$, giving support to our hypothesis of a monotonic relation between sHAR and sSFR. We then exploit the model to test a hypothesis on morphological evolution of galaxies. We attempt to explain the bulge/disk bimodality in terms of the two halo accretion modes: fast and slow accretion. Specifically, we speculate that bulge/spheroidal components might form during the early phase of fast halo growth, while disks form during the later phase of slow accretion. We find excellent agreement with both the observational bulge and elliptical mass functions.
\end{abstract}

\keywords{galaxies: formation - galaxies: evolution - galaxies: haloes - galaxy: morphology}

\section{Introduction}\label{sec:introduction}
In recent years, significant progresses have been made in our understanding of the processes leading to the formation and evolution of galaxies. This is largely due to increasingly sophisticated facilities that allow us to observe galaxies at higher redshifts and in different bands and to new theoretical models allowing to interpret these observations in a coherent framework. Despite numerous advancements, many crucial aspects of galaxy formation and evolution remain hotly debated and still unsolved.

For example, it is well established that dark matter (DM) structures are formed in a hierarchical way, with larger haloes gradually forming out of the merging of smaller ones and/or mass accretion from filaments. However, observations suggest that star formation and stellar mass assembly tend to follow a downsizing trend: massive galaxies form a significant part of their stellar mass earlier and rapidly, with a burst of intense star formation, and smaller ones form later and over longer timescales (see e.g. Cowie et al. 1996; Thomas et al. 2005, 2010; Merlin et al. 2019; Lah et al. 2022; Nanayakkara et al. 2022). Explaining these opposite trends is particularly challenging for current models, which tend to accurately predict the properties and the relations of local galaxies, but struggle in reproducing the strong star formation episodes observed at high redshift (see Fontanot et al. 2007; Somerville et al. 2012; Somerville \& Dav\'e 2015; Hirschmann et al. 2016; Bassini et al. 2020; Lovell et al. 2021; Hayward et al. 2021; Lustig et al. 2023; Dome et al. 2023), unless ad-hoc model assumptions are invoked, such as a non universal initial mass function (Baugh et al. 2005; Lacey et al. 2016; Fontanot et al. 2017). As a consequence, physical processes regulating star formation and quenching at high redshift are still poorly understood.

Cosmological hydrodynamic simulations (e.g. Vogelsberger et al. 2014; Hopkins et al. 2014; Crain et al. 2015; Schaye et al. 2015; Pillepich et al. 2018a, 2018b; Dav\'e et al. 2019) and semi analytic models (e.g. Lacey et al. 2016; Hirschmann et al. 2016; Lagos et al. 2018; Henriques et al. 2020) are at present the most comprehensive approaches at our disposal to probe the fine details of the origin and evolution of galaxies. Given the complexity of the multitude of processes at work, however, it is inevitable that multiple assumptions and associated parameters are adopted in input in semi analytic models and in the sub-grid, unresolved processes of simulations. It is thus expected, and in fact witnessed in several comparison works (e.g., Scannapieco et al. 2012), that degeneracies could occur, when similar observables can be reproduced by distinct models, limiting our baseline understanding of the actual physics of galaxy formation. A third complementary approach to probe galaxy evolution is semi empirical models, put forward in more recent years by different groups (Hopkins et al. 2009; Moster et al. 2013, 2018; Behroozi et al. 2013, 2019; Shankar et al. 2014; Mancuso et al. 2016; Buchan \& Shankar 2016; Grylls et al. 2019; Fu et al. 2022). Semi empirical models do not attempt to model the physics regulating the baryon cycle from first principles, but marginalize over it exploiting existing empirical relations between galaxies and DM haloes. By design, being data-driven, semi empirical models are characterized by a fewer number of assumptions and parameters than the other two modelling techniques. However, since those parameters are directly connected to observations and not to specific physical processes, making inference about physics on the basis of semi empirical models  is challenging and their scope is generally less ambitious than numerical simulations or semi analytic models. On the other hand, semi empirical models are useful in at least two respects: first, by empirically linking different observables, they can test for possible inconsistencies among distinct data sets, which can often occur given the significant systematics in, e.g., stellar mass or star formation rate measurements; second, new semi empirical models are expressly designed around minimal input assumptions and associated parameters; additional hypotheses can be gradually included in a bottom-up approach, one by one.

Starting from a DM merger tree, semi empirical models populate haloes by matching statistical distributions of 2 different quantities, one related to DM haloes and one to the baryonic component. The very first attempts to link galaxies to host haloes were simply based on abundance matching between the luminosity function and halo mass function (e.g., Vale \& Ostriker 2004, 2006; Shankar et al. 2006). Grylls et al. (2019) and Fu et al. (2022) also used abundance matching in more recent years, but applied more widely at different redshifts, to retrieve the link between stellar mass growth, star formation, and mergers. Moster et al. (2018) and Behroozi et al. (2019), instead, directly parameterised the link between star formation rate (SFR) and the circular velocity of the host DM halo, and, by integrating in time the SFR along the halo tracks, along with mergers, found the best-fit solutions to the observed SFR distributions and galaxy stellar mass functions. One of the advantages of the latter approach is that the stellar mass-halo mass relation (SMHM) is predicted, along with the scatter around it, and separated for star-forming and quenched galaxies. 

In this paper we present a new semi empirical model based on the abundance matching between specific halo accretion rate (sHAR) and specific star formation rate (sSFR) of galaxies, i.e., we assume that galaxies with larger sSFR reside in haloes with larger sHAR. This is the main assumption of the model, introducing a free parameter which is the scatter on the sSFR-sHAR relation. The approach taken here builds on the previous semi empirical models in two respects: 1) it uses the basic idea of abundance matching, which has the advantage of limiting the number of free parameters describing the connection between galaxies and haloes, to effectively only one parameter, the intrinsic scatter in the (monotonic) correlation, and 2) it makes use of sSFR rather then integrated stellar mass. We apply the sSFR-sHAR correlation only to star-forming galaxies; quiescent ones are not required to follow it. This is due to the fact that physical causes of quenching are still not well understood and could be linked to baryonic processes, not only to halo properties. Therefore we prefer to be as agnostic as possible and not to impose quenched galaxies to reside in low sHAR haloes. Despite this, we do keep track of passive galaxies in a fully empirical way, by following the evolution of the quiescent galaxy stellar mass function at various redshifts and assuming a selection criterion for quiescent galaxies (see section \ref{sec:quenching} for more details). This selection criterion is the second assumption of our model, but, as we will explain in section \ref{sec:quenching}, it will not severely affect our results.

We initialize our galaxies at $z=0$ by populating the DM haloes following the SMHM. This represents our initial condition and introduces the second parameter of the model: the scatter around the SMHM. Using the assumption of sSFR-sHAR correlation, we are able to track galaxy evolution backward in cosmic time, allowing us to predict the stellar mass function and the SMHM at higher redshift, along with its associated scatter. This approach avoids making assumptions about the SMF and SMHM a priori, and instead provides real predictions for these quantities, except at z=0 where the SMHM and SMF represent our initial condition. We assess the validity of this approach by comparing our derived SMF to the observational determination by Davidzon et al. (2017) from the COSMOS2015 catalogue. The model strength is that it relies only on two hypotheses: the monotonicity between sSFR and sHAR and the selection criterion for quiescent galaxies, and one initial condition, i.e., the initialization of galaxies on the SMHM relation at $z=0$.

On the assumption of a monotonic relation between sSFR and sHAR, we are able to build at any given epoch a catalogue of mock galaxies with defined stellar masses and host halo masses. In the second part of this work we will adopt our mock galaxies and their assembly histories to predict their morphological appearance, more specifically the relative fraction of bulge-to-disk stellar masses. To achieve this goal, we will assume a two-phase galaxy formation scenario, linked to the two-mode DM halo accretion histories. Specifically, N-body simulations analyses (Wechsler et al. 2002; Zhao et al. 2003, 2009; Tasitsiomi et al. 2004; Diemand et al. 2007; Hoffman et al. 2007; Ascasibar \& Gottloeber 2008; More et al. 2015; Hearin et al. 2021) have highlighted that mass accretion history of haloes can be roughly divided into two distinct phases: an early fast accretion phase dominated by major mergers and violent collapse, which shapes the structure of the inner halo potential well, and a later slow accretion phase characterized by a smoother DM accretion or minor mergers and usually dominated by pseudo-evolution (Diemand et al. 2007; Diemer et al. 2013; Zemp et al. 2014; More et al. 2015), which contributes to growing the halo outskirts not altering the central structure. We suggest that bulges and spheroidal components might be formed during the early fast accretion phase, when violent accretion processes and dynamical friction between giant gas clumps may lead to a quick loss of angular momentum also for baryonic matter that can fastly sink to the very central region (Lapi et al. 2018a; Pantoni et al. 2019). Galactic stellar disks are instead formed during the later slow accretion phase when baryons can retain part of their angular momentum and are not directly funneled toward the center. Such an idea builds upon the pioneering works of Mo \& Mao (2004) and Cook et al. (2009) and can qualitatively explain stellar archaeological findings showing stellar population of bulges/spheroids to be older, $\alpha-$enhanced and rapidly generated, and stars in disks to be younger and formed over longer timescales (see e.g. Chiappini et al. 1997; Thomas et al. 2005, 2010; Gallazzi et al. 2006; Johansson et al. 2012a; Courteau et al. 2014; Pezzulli \& Fraternali 2016; Grisoni et al. 2017; Bellstedt et al. 2023). In the present work, we extend our simple semi empirical model to assess the validity of this idea with a very simple assumption: all stars formed during fast DM accretion are considered to be in the bulge, while those formed during slow accretion constitute the stellar disk. The interplay between the transition time from fast to slow accretion and the quenching time would lead to the generation of different types of galaxies, from disk-only galaxies to pure ellipticals. The model naturally predicts the stellar mass function for bulges and ellipticals and the fraction of ellipticals that we compare with the observational determination from GAMA survey by Moffett et al. (2016a, 2016b).

The paper is structured as follows. In section \ref{sec:haloes}, we build our catalogue of DM haloes using the prescriptions of Hearin et al. (2021) which have proven to well reproduce results from N-body simulations keeping into account, for a given descendant halo mass, not only the average halo growth but also the possible variance. In section \ref{sec:galaxies}, we present the main empirical recipes adopted for galaxies, such as the stellar mass function for star-forming and quenched galaxies and the main sequence. In section \ref{sec:empirical_model}, we describe the semi empirical model, discussing all the assumptions and free parameters; while in section \ref{sec:results}, we show the main results. In section \ref{sec:bulge_disk}, we refine the model introducing the hypothesis for morphological evolution and we compare our findings with observations. Finally, in section \ref{sec:conclusion}, we draw our conclusions.

\section{Dark Matter Haloes}\label{sec:haloes}
Any semi empirical model of galaxy evolution should lay its foundations on DM halo accretion histories. Since we want a discrete model following the evolution of single galaxies, we need to build a discrete catalogue of haloes. This section outlines the process of building such a catalogue and deriving the key statistical quantities needed for the model.

\subsection{Building the Catalogue}
Halo statistics and their evolution are completely set by N-body simulation results. In this work, we use the halo mass function (HMF) determination from Tinker et al. (2008), which accounts for isolated haloes and central haloes in groups/clusters. This means our model traces field galaxies and central galaxies in groups/clusters, not taking into account satellites. We create a mock catalogue of $N_{\rm halo}=5\times 10^5$ haloes by sampling the HMF at $z=0$ for $\log M_H>10.75$. We choose this lower limit for 2 main reasons: (i) given the total number of simulated haloes, it guarantees enough resolution to follow the evolution of more massive ones, (ii) it ensures that the contribution of satellite haloes to the HMF is subdominant (see e.g. Aversa et al. 2015; Behroozi et al. 2019; Ronconi et al. 2020; Fu et al. 2022). We evolve our mock catalogue backwards in time, taking into account not only the average halo accretion history for a given descendant mass, but also the variance between different haloes, thus producing an accretion history halo-by-halo. To do this in a fast and flexible way we exploit the \texttt{DIFFMAH} code from Hearin et al. (2021) which well reproduces N-body simulations both for the average halo accretion history and for its scatter, randomly producing a variety of mass accretion histories for a given descendant halo mass. After these steps, we are left with a mock catalogue of haloes, each one with its own realistic history $\{M_{H,i}(z)\}$.

We then divide the accretion history of each halo into two modes: fast accretion and slow accretion. We say a halo is in the fast accretion regime if $\dot M_H/M_H>\Gamma\,H(z)$ with $\Gamma=1.5$ (see More et al. 2015); it usually occurs at high redshift and represents the phase in which the halo is able to accrete matter at a very fast rate, via major mergers or strong inflows, constituting the bulk of its potential well. An halo is in slow accretion regime if $\dot M_H/M_H<\Gamma\,H(z)$, meaning that matter 
is accreted at a much slower rate and possibly infuenced by pseudo-evolution, i.e., a spurious evolution of the virial mass of the DM halo due to the decrease of the background density with cosmic time, but not corresponding to an actual accretion of matter (see Diemand et al. 2007; Diemer et al. 2013, Zemp et al. 2014, More et al 2015). We label the transition redshift between fast and slow accretion for any given halo as $z_{\rm FS}$. N-body simulations analyses (Wechsler et al. 2002; Zhao et al. 2003, 2009; Tasitsiomi et al. 2004; Diemand et al. 2007; Hoffman et al. 2007; Ascasibar \& Gottloeber 2008; More et al. 2015) have revealed profound differences between these two accretion phases. In particular, during fast accretion, the circular velocity of the halo tends to rapidly increase and rapid variations in the inner profile occur, with $r_s$ and $M(<r_s)$ increasing at fixed concentration $c\sim 4$. This indicates that the central halo potential well is shaped during this phase. On the other hand, during slow accretion, the inner circular velocity $v(r_s)$ is approximately equal to the circular velocity in the fast accretion phase, the radius $r_s$ and $M(<r_s)$ tend to remain constant, while the virial radius and mass increase. As a consequence, the concentration parameter increases as well, indicating that the newly accreted matter is not perturbing the central potential but it is contributing to build the halo outskirt. We make use of these distinct behaviors of DM accretion in section \ref{sec:bulge_disk} to differentiate between the formation of bulges and disks in galaxies. 

\subsection{Halo Mass Function}
The halo mass function for our catalogue ($\rm HMF_{\rm cat}$) at different redshifts can be computed as:

\begin{equation}
\begin{split}
\rm HMF_{\rm cat}\equiv&\frac{\rm dN_{\rm cat}}{dV\,d\log M_H}=\\
&=\frac{1}{\rm N_{\rm halo}}\,\frac{\rm dN}{\rm dV}\,\sum_i{\delta(\log M_H-\log M_{H,i})},
\end{split}
\label{eq:HMF_cat}
\end{equation}
where $M_{H,i}$ is the mass of the $i-$th halo at redshift $z$ and the factor $\rm d^2N/dV/N_{\rm halo}$ is just a normalization to account for the correct halo number density.

\begin{figure*}
\centering
\includegraphics[width=0.75\textwidth]{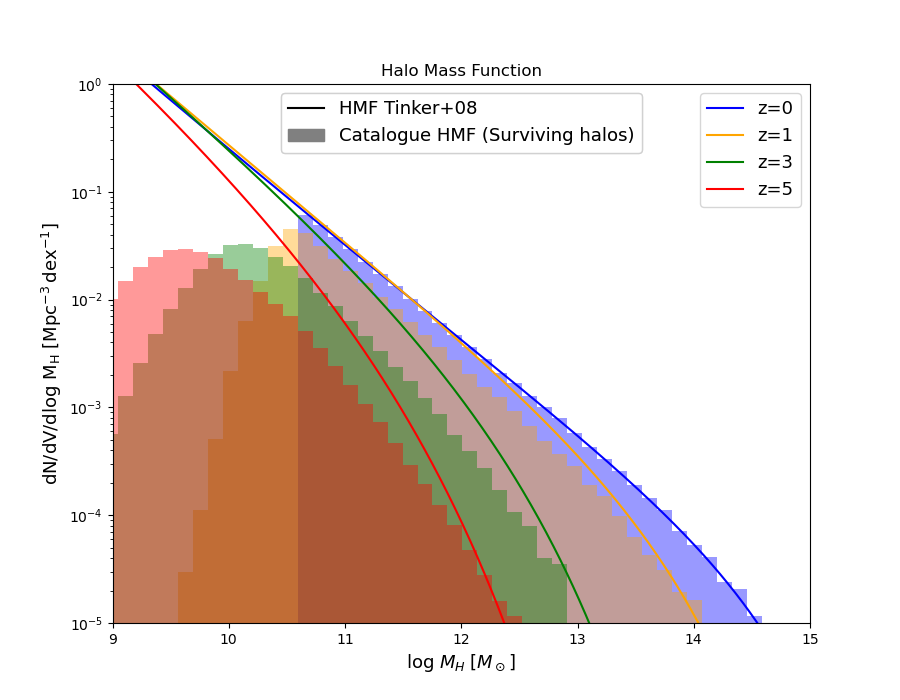}
\caption[]{Halo mass function at different redshifts: blue $z=0$, orange $z=1$, green $z=3$, and red $z=5$. Histograms represent the HMF of our mock catalogue of surviving haloes, solid lines are the HMF determinations by Tinker et al. (2008). The difference between HMF for the catalogue and the Tinker et al. (2008) determination is due to the fact that the catalogue keeps into account only surviving haloes at $z=0$.}
\label{fig:HMF}
\end{figure*}

Figure \ref{fig:HMF} shows the comparison between the catalogue $\rm HMF_{\rm cat}$ and the true HMF from Tinker et al. (2008) ($\rm HMF_{\rm true}$) at various redshifts. From Figure \ref{fig:HMF} we can see that our mock catalogue at $z>0$ well-reproduces the true HMF at the bright end, but it does not match the correct normalization at intermediate and low halo masses. The reason for this is that our catalogue only tracks the evolution histories of surviving haloes at $z=0$. They constitute just a small fraction of the total halo population at higher redshifts. The remaining haloes, not tracked by our catalogue, disappeared in time due to mergers with larger haloes and subsequent disruption, e.g. via tidal disruption events. We can define the quantity:

\begin{equation}
p(M_H,z)=\frac{\rm HMF_{\rm true}(M_H,z)}{\rm HMF_{\rm cat}(M_H,z)},
\end{equation}
as the ratio between the true and the surviving HMF obtained from the catalogue at each halo mass and redshift, and the quantity:

\begin{equation}
p_i=p_i(M_{H,i},z)=\frac{\rm HMF_{\rm true}(M_{H,i},z)}{\rm HMF_{\rm cat}(M_{H,i},z)},
\end{equation}
which represents the weight we should assign to each halo of the catalogue to reproduce the true number of haloes at that specific mass and redshift. 

The true HMF can be retrieved from the catalogue just by weighting the sum in equation \eqref{eq:HMF_cat} by the weight $p_i$ assigned to each halo:

\begin{equation}
\begin{split}
\rm HMF_{\rm true}\equiv&\frac{\rm d^2N_{\rm true}}{dV\,d\log M_H}=\\
&=\frac{1}{\rm N_{\rm halo}}\,\frac{\rm dN}{\rm dV}\,\sum_i{\delta(\log M_H-\log M_{H,i})\,p_i}.
\end{split}
\end{equation}
In the remainder of the work we will exploit $p_i$ to pass from catalogue statistics to true statistics.

\subsection{Specific Halo Accretion Rate Function}\label{sec:shar}
A useful quantity for the model we are going to build is the specific halo accretion rate, defined as: 

\begin{equation}
\rm sHAR\equiv\frac{\dot M_H}{M_H}.
\end{equation}
Its statistical distribution for the catalogue can be computed as:

\begin{equation}
\frac{\rm d^2N_{\rm cat}}{\rm dV\,d\log sHAR}=\frac{1}{\rm N_{\rm halo}}\,\frac{\rm dN}{\rm dV}\,\sum_i{\delta(\rm \log sHAR-\log sHAR_{i})},
\label{eq:SHARF}
\end{equation}
where $\rm sHAR_i$ is the specific star formation rate of the $i-$th halo at redshift $z$. The true sHAR function at given $z$ should also keep track of the contribution of non-surviving haloes, not present in the mock catalogue. The expression for the true halo specific accretion rate function is the following:

\begin{equation}
\begin{split}
\frac{\rm d^2N_{\rm true}}{\rm dV\,d\log sHAR}=&\frac{1}{\rm N_{\rm halo}}\,\frac{\rm dN}{\rm dV}\times\\
&\times\sum_i{\delta(\rm \log sHAR-\log sHAR_{i})\,p_i}.
\end{split}
\label{eq:dpdlogSHAR}
\end{equation}
The assumption under this equation is that, for a given $M_H$ and $z$, the sHAR distribution of surviving haloes is the same as the sHAR distribution of non-surviving haloes up to the moment at which they are accreted as satellites. \footnote{Note that a satellite could have a sHAR drastically different from a central or isolated halo, attaining negative sHAR values due to stripping. However, since our model adopts the Tinker et al. (2008) HMF, not including satellites, these haloes would disappear from the HMF once they are accreted. Therefore our assumption that surviving and non-surviving haloes feature the same sHAR distribution has to be valid only until the time at which satellites are accreted, not during their subsequent evolution as members of the cluster.}

\begin{figure*}
\centering
\includegraphics[width=0.75\textwidth]{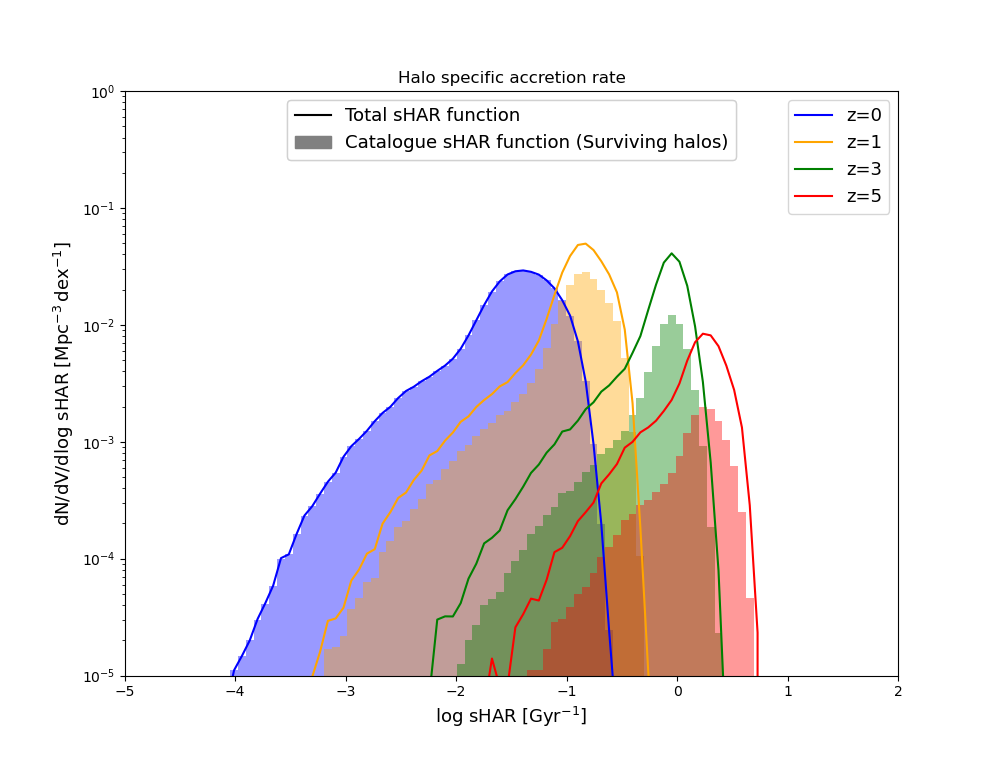}
\caption[]{sHAR functions. Colors are as in Figure \ref{fig:HMF}. Histograms are the sHAR distributions for our mock catalogue, solid lines the reconstructed total sHAR for all the haloes. sHAR tend to increase with redshift on average.}
\label{fig:SHARF}
\end{figure*}

Distributions in equations \eqref{eq:SHARF} and \eqref{eq:dpdlogSHAR} are shown in Figure \ref{fig:SHARF}, where different colors stand for different redshifts. Histograms represent the sHAR distribution for surviving haloes in our mock catalogue, while solid lines are the reconstruction of the overall sHAR for both surviving and destroyed haloes. We can see that the sHAR functions are peaked functions with average increasing at higher redshift. Since the sHAR functions are peaked, we can easily define the probability distribution of sHAR at fixed $z$ ($\rm dp/d\log sHAR$) just by normalizing them to unity. Such a distribution will be used in section \ref{sec:abundance_matching} to perform abundance matching and derive the sSFR-sHAR relation, once the contribution from quiescent galaxies has been removed (see section \ref{sec:quenching}). 

We stress that, even if in Figure \ref{fig:SHARF} we have plotted the sHAR distribution for all the haloes in our catalogue, an important result we have proven is that the sHAR distribution does not depend on the halo mass. Haloes in different mass bins feature the same sHAR distribution shown in Figure \ref{fig:SHARF}. This means that the probability distribution of having a certain sHAR does not depend on the mass of the halo:

\begin{equation}
\frac{\rm dp}{\rm d\log sHAR}(\rm sHAR,z|M_H)=\frac{\rm dp}{\rm d\log sHAR}(\rm sHAR,z).
\end{equation}
This result is important because it guarantees that the selection criterion for quenched galaxies does not affect the derived sSFR-sHAR relation (see the end of section \ref{sec:quenching} for more details), making our results robust against the second assumption of our model (i.e., the selection of quenched galaxies).

\section{Empirical Data for Galaxies}\label{sec:galaxies}
For the treatment of the baryonic component we rely on empirical relations at different redshifts, such as the stellar mass functions and the main sequence of star-forming galaxies, which give us a snapshot of galaxy properties at every cosmic time. Connecting these snapshots together is the aim of our semi empirical model.

\subsection{Stellar Mass Functions}
\begin{figure*}
\centering
\includegraphics[width=0.75\textwidth]{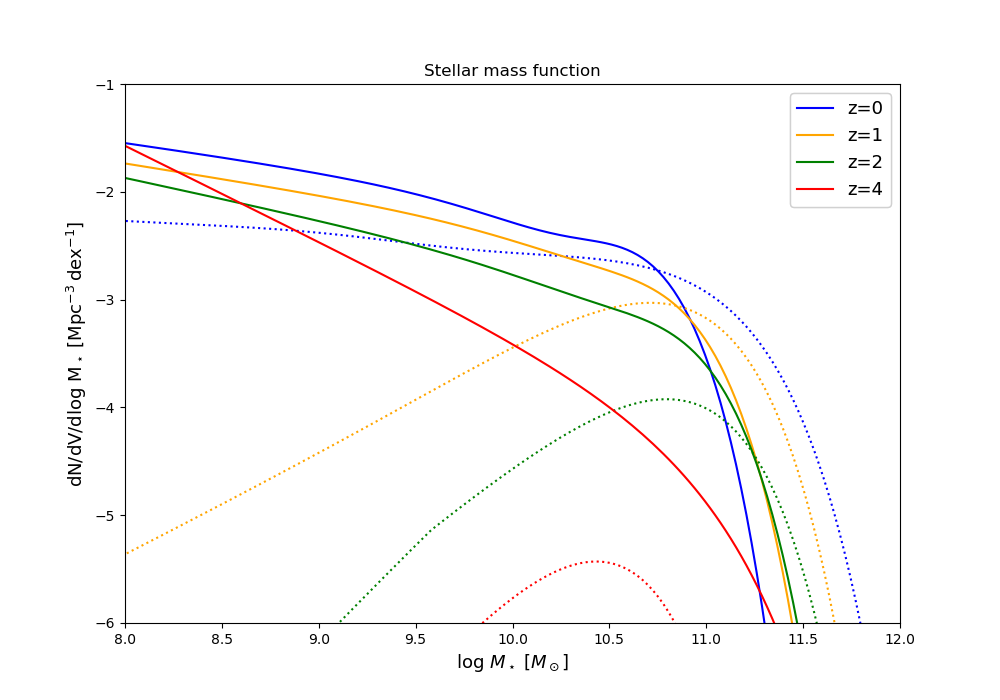}
\caption[]{SMF from Davidzon et al. (2017). Colors are as in Figure \ref{fig:HMF}. Solid lines are the SMF for star-forming galaxies, dashed lines for quenched.}
\label{fig:SMF}
\end{figure*}

Stellar mass functions are one of the key ingredients to investigate galaxy evolution. An unbiased calibration of the SMF at different redshifts could help in shedding light on the assembly histories of galaxies in a statistical sense. Given their importance, several observational estimates have been proposed in literature at various redshifts and stellar mass ranges (e.g. Ilbert et al. 2013; Moustakas et al. 2013; Muzzin et al. 2013; Bernardi et al. 2013, 2016, 2017; Davidzon et al. 2017). Stellar mass function determinations may differ from each other since they are usually obtained from SED fitting which can depend on the stellar population model, star formation history and IMF. Moreover different estimates are derived from various surveys possibly with different characteristics and observing different regions of the sky. For the sake of simplicity, we adopt the SMF by Davidzon et al. (2017) for both star-forming and quenched galaxies ($\rm dN_{\rm ac}/dV/d\log M_\star$ and $\rm dN_{\rm pas}/dV/d\log M_\star$ respectively). The fits proposed by Davidzon et al. (2017) are shown in Figure \ref{fig:SMF} (solid lines for star-forming galaxies and dashed lines for quenched). From the SMF of active and quenched galaxies we can derive the fraction of quenched galaxies for each stellar mass and redshift:

\begin{equation}
f_Q(M_\star,z)=\frac{\rm SMF_{\rm pas}}{\rm SMF}.
\label{eq:fQ}
\end{equation}

\subsection{Main Sequence}
In order to determine the SFR $\psi$ distributions competing to star-forming galaxies at any given epoch, we consider the main sequence of star-forming galaxies, a well-known correlation between stellar mass and SFR at given redshift. Analogously to the SMF, even for the main sequence, many groups have attempted to define, both observationally and theoretically, its precise form across cosmic time  (see, e.g., Daddi et al. 2007; Rodighiero et al. 2011, 2015; Speagle et al. 2014; Whitaker et al. 2014; Schreiber et al. 2015; Mancuso et al. 2016b; Dunlop et al. 2017; Bisigello et al. 2018; Pantoni et al. 2019; Lapi et al. 2020; Popesso et al. 2023; Huang et al. 2023). Despite all these efforts, its shape, redshift evolution and scatter are still debated, especially at the high mass end. In the present work, we choose to use the recent determination by Popesso et al. (2023), which is a compilation of many literature studies, converted to a common calibration, over wide redshift and stellar mass ranges. However, the main sequence is only an average relation in the $\rm M_\star-SFR$ plane, featuring a certain degree of dispersion and significant outliers. Many studies (e.g. Bethermin et al. 2012; Sargent et al. 2012; Ilbert et al. 2015; Schreiber et al. 2015) have highlighted that, at fixed stellar mass and redshift, star-forming galaxies tend to be distributed in SFR according to a double Gaussian shape, reflecting the well-known galaxy bimodality between main sequence and starburst galaxies. In this work, we model this distribution as in Sargent et al. (2012):

\begin{equation}
\begin{split}
\frac{\rm dp}{\rm d\log\psi}(\psi|z,M_\star)&=\frac{A_{\rm MS}}{\sqrt{2\pi\,\sigma_{\rm MS}^2}}\exp{\left[-\frac{(\log\psi-\langle\log\psi\rangle_{\rm MS})^2}{2\sigma_{\rm MS}^2}\right]}+\\
&+\frac{A_{\rm SB}}{\sqrt{2\pi\,\sigma_{\rm SB}^2}}\exp{\left[-\frac{(\log\psi-\langle\log\psi\rangle_{\rm SB})^2}{2\sigma_{\rm SB}^2}\right]},
\end{split}
\label{eq:dpdlogpsi}
\end{equation}
where $A_{\rm MS}=0.97$ is the fraction of main sequence galaxies,  $A_{\rm SB}=0.03$ the fraction of starbursts, $\langle\log\psi\rangle_{\rm MS}$ the value given by the main sequence and representing the central value for the first Gaussian, $\langle\log\psi\rangle_{\rm SB}=\langle\log\psi\rangle_{\rm MS}+0.59\,\rm dex$ the central value of the second Gaussian, $\sigma_{\rm MS}=0.188\,\rm dex$ the one-sigma dispersion of the first Gaussian and $\sigma_{\rm SB}=0.243\,\rm dex$ the dispersion of the starburst population.

\begin{figure*}
\centering
\includegraphics[width=0.75\textwidth]{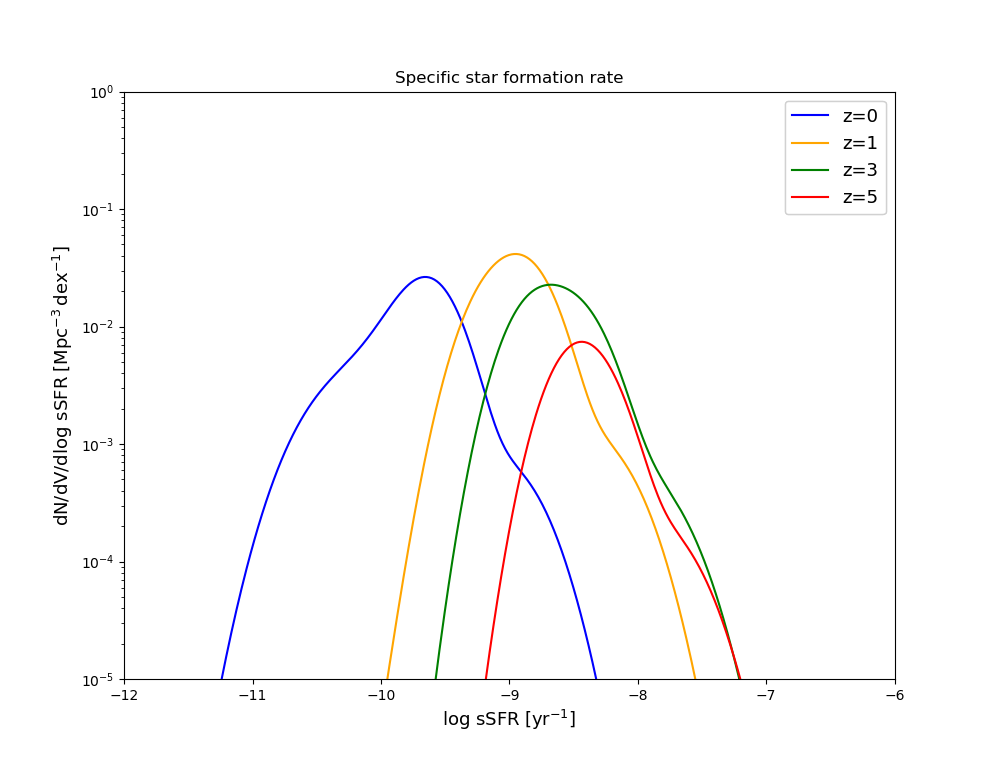}
\caption[]{sSFR distribution from the convolution of the SMF of active galaxies and the main sequence. Colors are as in Figure \ref{fig:HMF}.}
\label{fig:dotNdVdsSFR}
\end{figure*}

The specific star formation rate, which is the quantity of interest in this work, is defined as: 

\begin{equation}
    \rm sSFR\equiv\frac{\psi}{M_\star},
\end{equation}
and the sSFR functions $\rm d^2N/dV/d\log sSFR$ can be computed as:

\begin{equation}
\begin{split}
    \frac{\rm d^2N}{\rm dV\,d\log sSFR}&(\rm z,\log sSFR)=\\
    &=\int\rm d\log M_\star\,\frac{\rm d^2N_{\rm ac}}{\rm dV\,d\log M_\star}(z,\log M_\star)\times\\
    &\times\frac{\rm dp}{\rm d\log\psi}(\log\psi=\log sSFR+\log M_\star|z,M_\star).
    \end{split}
\label{eq:dotNdVdsSFR}
\end{equation}
This last quantity is shown in Figure \ref{fig:dotNdVdsSFR}, with different colors standing for different redshifts. We can see that the sSFR functions are characterized by a main peak representing main sequence galaxies and a secondary peak at higher sSFR representing starbursts. The average sSFR increases with redshift. Since sSFR functions are peaked, as in the case of sHAR, we can easily compute the sSFR probability distribution at different redshifts ($\rm dp/d\log sSFR$), normalizing them to unity.

\section{The Empirical Model}\label{sec:empirical_model}
In the previous sections we have described the main ingredients needed to construct our semi empirical model: (i) the results of N-body simulations for the halo part and (ii) empirical relations such as the SMF and the main sequence for the baryonic component. We want now to connect DM to baryons, populating haloes with stars in an empirical approach. One of the traditional choices to map baryons to haloes is via abundance matching, which consists in assuming a monotonic relationship between two quantities, $X$ and $Y$, one related to haloes and one to baryons, and matching their statistical distributions. The problem main problem of abundance matching is that perfect monotonic relations do not exist: galaxies and haloes will always feature a certain distribution in the $X-Y$ plane, given by the degeneracies with many other nuisance parameters. Abundance matching techniques do not allow to derive the full distribution in the $X-Y$ plane but just an average relation and, even when scatter is kept into account, it has to be set a priori (see Aversa et al. 2015). For this reason, every semi empirical model based on abundance matching requires an assumption of monotonicity between $X$ and $Y$ and a free parameter which is the scatter on the $X-Y$ relation. On the other hand, abundance matching techniques, taking as input these limited assumptions, are a powerful tool to understand how $X$ and $Y$ relate to other observables and to derive the relations and full distributions between all the other relevant quantities. 

Plainly, the variables $X$ and $Y$ have to be accurately chosen, by selecting 2 quantities we believe to be in a tight relationship. The usual choice in this respect is to perform abundance matching between stellar mass and halo mass, assuming a monotonic SMHM relation with scatter inserted a priori. In this paper, instead, we choose to perform abundance matching between the specific halo accretion rate and the specific star formation rate for galaxies. This assumption is guided by 2 main ideas. First of all, while stellar and halo masses are integrated quantities, which means they can depend on the history of mass accretion, we introduce differential quantities in the abundance matching, i.e., star formation rate and halo accretion rate (HAR), better capturing the situation at that specific time. We expect the SFR of a galaxy to be somewhat linked to the HAR since a faster DM accretion would correspond to a faster gas inflow and, consequently, to more gas available to form stars. On the other hand, star formation efficiency, the amount of baryons converted into stars, can heavily change for objects with different masses (see e.g. Moster et al. 2010, 2013 2018; ; Behroozi et al. 2013, 2019; Aversa et al. 2015; Rodriguez-Puebla et al. 2015), possibly creating some scatter around the mean SFR-HAR relation. To marginalize over this effect, we decide to perform abundance matching between two mixed quantities sSFR and sHAR which are the ratios between the derivative of the mass and the mass itself at a given time:

\begin{align}
    \frac{\psi}{M_\star}=f\left(\frac{\dot M_H}{M_H}\right),
    \label{eq:abma_s}
\end{align}

where $f(\dot M_H/M_H)$ is a monotonic function to be found by abundance matching. Therefore the main assumption in our model is the monotonicity in the sSFR-sHAR relation and the first free parameter is the scatter $\sigma_{\rm sSFR}$ around this relation. As we will see in the remaining part of this section, the latter assumption will provide a way to populate haloes with stars and to obtain a mock catalogue of galaxies, each with its own star formation history (SFH). Using this catalogue we can then derive other relevant quantities, such as the SMF and the distribution in the $M_H-M_\star$ plane, which become predictions of the model; we show them in section \ref{sec:results}. 

At this stage, two important remarks are in order. First of all, we assume that only star-forming galaxies follow the sSFR-sHAR relation, not quenched ones. Indeed, since equation \eqref{eq:abma_s} connects sSFR to the DM halo properties, if applied to all the haloes and galaxies, it would naturally provide an explanation for quenching related to the accretion history of the DM halo: quenched galaxies, having low sSFR, would be classified as the ones embedded in low sHAR haloes. However, the physical reasons of quenching are still largely debated and possibly related to baryon physics, such as some form of feedback connected to the growth of a supermassive black hole (SMBH) in the center of the galaxy. Since the aim of this work is not to pin down the physical mechanism leading to quenching, we try to be as agnostic as possible in regards to this problem. For this reason we decide to treat quenched galaxies in a fully empirical way (see section \ref{sec:quenching}) and we assume equation \eqref{eq:abma_s} to be valid only up to the moment of quenching; after that time stellar mass does not grow anymore, independently of the accretion history of the DM halo.

Secondly, sSFR alone is not useful to fully characterize a galaxy star formation history if we do not know the stellar mass at a certain time (translating sSFR to SFR requires the knowledge of $M_\star$). For this reason we need an initial condition for the stellar masses of our mock galaxies. In the next sections we discuss the initialization of our galaxies (section \ref{sec:initialization}) and the treatment of quenched galaxies (section \ref{sec:quenching}).

\subsection{Initialization}\label{sec:initialization}
As stated above, in order to derive a SFH from the sSFR we need to know the stellar mass at a certain initial time. We decide to initialize our galaxy masses at $z=0$, since we dispose of more complete and robust observational data in the local Universe with respect to high redshift, and follow their SFH backwards in time. Such initialization is done on the basis of the $z=0$ SMHM. This relation is obtained via abundance matching of stellar masses with halo masses, but its validity has also been confirmed by lensing data (see Reyes et al. 2012; Mandelbaum et al. 2016; Lapi et al. 2018b). The $M_\star(M_H,z)$ relation is used at $z=0$ to initialize our catalogue with stellar masses, i.e., $M_{\star,i}(z=0)=M_\star(M_{H,i},z=0)$. The scatter on this relation $\sigma_{M_\star}$ is the second parameter of the model. We stress that, while we use the $z=0$ SMHM as initial condition, the evolution of the relation is a prediction of our model. 

\subsection{Treatment of Quenched Galaxies}\label{sec:quenching}
A key element of the evolutionary history of galaxies is quenching: some galaxies halt or strongly diminish their star formation activity at a certain time because of some physical mechanisms possibly due to halo/environmental conditions, such as starvation, strangulation or shock heating (Birnboim \& Dekel 2003; Keres et al. 2005; Dekel \& Birnboim 2006; Dekel et al. 2009) or feedback processes, e.g. supernova or AGN feedback (see Silk \& Rees 1998; King 2003, 2014; Granato et al. 2004; Di Matteo et al. 2005; Lapi et al. 2006, 2014; Shankar et al. 2006; Fabian 2012; King \& Pounds 2015; Mancuso et al. 2016; Ma et al. 2022; Bluck et al. 2023). The aim of this work is not to pin down the physical causes behind quenching, but to simply empirically keep track of quenched galaxies. 

We start by selecting quiescent galaxies at $z=0$ and then tracing their evolution backwards in time. To achieve these two steps we make use of two important quantities: (i) $f_Q(M_\star,z)$, defined in equation \eqref{eq:fQ}, labelling the fraction of quenched galaxies with respect to the total as a function of stellar mass and redshift, and (ii) the ratio between the number density of passive galaxies at 2 different redshifts for a given stellar mass bin:

\begin{equation}
f_{Q-Q}(M_\star,z_1,z_2)=\frac{\rm SMF_{\rm pas}(z_2,M_\star)}{\rm SMF_{\rm pas}(z_1,M_\star)},
\label{eq:fQ_Q}
\end{equation}
with $z_1<z_2$. Equation \eqref{eq:fQ_Q} provides the fraction of quiescent galaxies of stellar mass $M_\star$ present at $z_1$, that were already quiescent at the higher redshift $z_2$. Correspondingly, its reciprocus $1-f_{Q-Q}$ details the fraction of quiescent galaxies at $z_1$ that were star-forming at $z_2$. The latter parameter therefore allows to trace the behavior of quenched objects at different redshifts.

Given the definitions of the fractions of quenched/star-forming galaxies at each time step and stellar mass, we can now proceed in including quenching into our model. Below we list the steps we follow to calculate the fraction of quenched galaxies:
\begin{enumerate}
\item We start at $z=0$ where we associate a stellar mass $M_{\star,i}$ to each halo of the catalogue with halo mass $M_{H,i}$ using the SMHM at $z=0$.
\item We use the factor $f_Q(M_{\star},z=0)$ to distinguish the fractions of quiescent and star-forming galaxies of stellar mass $M_\star$ at z=0.
\item We slightly increase redshift (from $z$ to $z+dz$) and we use $1-f_{Q-Q}(M_{\star},z,z+dz)$ to predict the fraction of galaxies with stellar mass $M_\star$ that become active between $z$ and $z+dz$.
\item We repeat step 3 up to high redshifts.
\end{enumerate}
Before performing these steps it is important to discuss two caveats to this methodology that I present concisely below and discuss extensively in Sections \ref{subsec:modification} and \ref{subsec:selection_criterion}: 
\begin{itemize}
\item We need to slightly modify the coefficient in equation \eqref{eq:fQ_Q} to be used in our mock catalogue. We discuss this in section \ref{subsec:modification}.
\item We need a criterion to select which galaxies with given $M_\star$ are chosen as quiescent and which as star-forming. We discuss the choice of this criterion in section \ref{subsec:selection_criterion}.
\end{itemize}

\subsubsection{Modifications to $f_{Q-Q}$}\label{subsec:modification}
In the computation of the factor $f_{Q-Q}$ (equation \eqref{eq:fQ_Q}) we are using the true stellar mass functions, i.e., the ones derived by the COSMOS2015 catalogue, for all the galaxies present at a given redshift. From the halo treatment, we know that our catalogue features an increasing lack of haloes (and consequently of galaxies) going towards higher redshift, since it keeps track of only surviving haloes (section \ref{sec:haloes}). Therefore, the ratio between the numbers of quenched galaxies at different $z$ for our catalogue will be influenced by the weight $p(M_H,z)$. In particular:

\begin{equation}
\begin{split}
f_{Q-Q, \rm cat}&(M_\star,M_H,z_1,z_2)=\frac{\rm SMF_{\rm cat,pas}(z_2,M_\star)}{\rm SMF_{\rm cat,pas}(z_1,M_\star)}=\\
&=\frac{\rm SMF_{\rm pas}(z_2,M_\star)}{\rm SMF_{\rm pas}(z_1,M_\star)}\,\frac{p(z_1,M_H)}{p(z_2,M_H)}.
\end{split}
\label{eq:fQ_Q_cat}
\end{equation}
This $f_{Q-Q, \rm cat}(M_\star,M_H,z_1,z_2)$ is the factor we have to use for our mock catalogue. To gain a better physical grasp of the meaning of this let us make an example with real numbers. Let us fix the stellar mass $M_\star$ and let us assume the factor $f_{Q-Q}$ between $z$ and $z+dz$ is $f_{Q-Q}(M_\star,z,z+dz)=0.9$. This means that $90\%$ of quenched galaxies at the lower redshift $z$ were already quenched at $z+dz$. The remaining $10\%$ will become star-forming going from $z$ to $z+dz$\footnote{We remind the reader that we are looking at galaxy evolution backwards in time. So whenever we say galaxies "becomes" star-forming or are "activated" we are not meaning reactivation or rejuvenation of quenched galaxies, but we are simply evolving the galaxy backwards until it reaches a redshift were it was star-forming. We also remind that in this work, the words "active" and "passive" have nothing to do with AGN activity, but they are just synonyms of star-forming and quiescent, respectively.}. This is valid for real galaxies. However, in our mock Universe, as stated above, the number of galaxies with given mass decreases at higher redshifts with respect to the real case. Therefore, if in the catalogue we have $100$ quenched galaxies at $z$, we cannot keep $90$ quenched galaxies at $z+dz$ because in this way we would automatically impose that the overall difference between the true and the mock number of galaxies affects only star-forming objects. Instead we need to reduce the number of quenched galaxies according to the ratio $p(z,M_H)/p(z+dz,M_H)$.

\begin{figure*}
\centering
\includegraphics[scale=0.7]{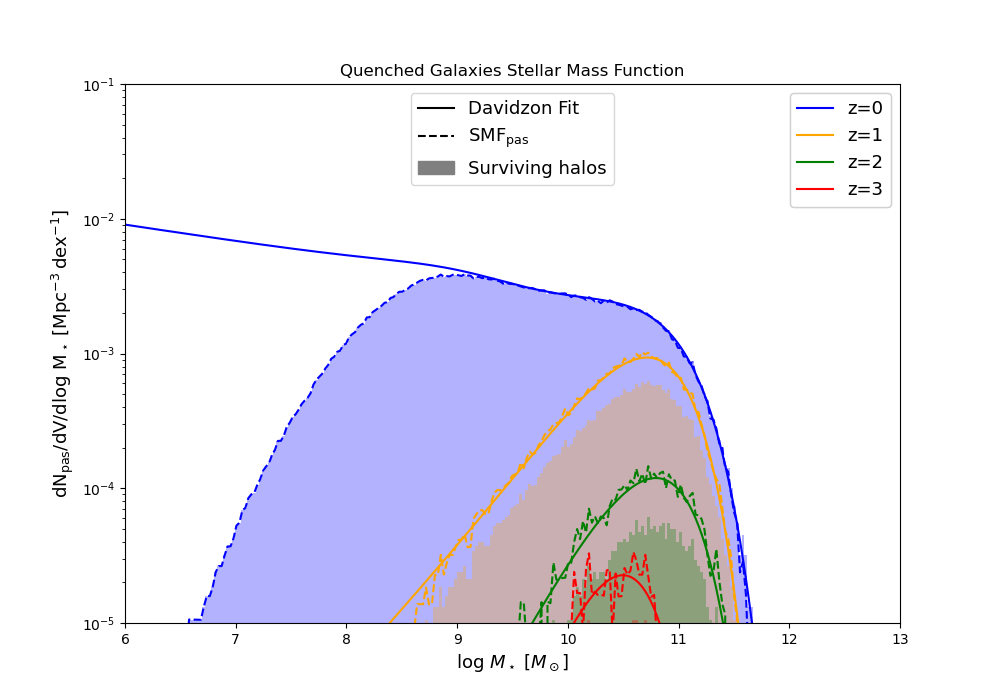}
\caption[]{SMF for quenched galaxies. Colors are as in Figure \ref{fig:HMF}. Histograms are the SMF for surviving haloes, dashed lines the total SMF reconstructed from our catalogue and solid lines the fit by Davidzon et al. (2017). By contruction, our model tracks the evolution of the quenched galaxies SMF.}
\label{fig:SMF_quenched_distance}
\end{figure*}

Once this correction is taken into account, we can use the steps described in section \ref{sec:quenching} to trace the evolution of quiescent galaxies backwards in time. In Figure \ref{fig:SMF_quenched_distance} we show the mass function for quenched galaxies at different redshifts (color code). Solid lines are the fit by Davidzon et al. (2017), histograms represent the distribution of mass for quenched galaxies in our mock catalogue (obtained by the use of the factor $f_{Q-Q, \rm cat}$), while dashed lines the reconstruction of the passive SMF obtained for all galaxies by weighting galaxies in our catalogue with the factor $p_i$. From the Figure it is evident that our method is able to reproduce the SMF of quenched galaxies at all redshifts, as expected by construction, testifying the validity of the approach outlined here to track the evolution of quiescent galaxies.

\subsubsection{Selection Criteria for Quenched Galaxies}\label{subsec:selection_criterion}
Factors as $f_Q$ and $f_{Q-Q,\rm cat}$ allow us to empirically follow the evolution of quenched galaxies, indicating, at each redshift and in each stellar mass bin, how many galaxies should be kept quenched and how many should be activated. The results are correct, as shown in Figure \ref{fig:SMF_quenched_distance}. However, there is still one point to consider in the selection of quenched galaxies in the mock. At fixed redshift and stellar mass, how should we select the galaxies we are going to keep quenched and the ones we are going to activate? At random or following some criteria based on halo mass or sSFR threshold? At some level, the criteria are somewhat arbitrary. Notice that any possible choice would not affect the SMF for quenched objects (Figure \ref{fig:SMF_quenched_distance}), which only represents the number of quenched galaxies at each epoch in each stellar mass bin, but it would affect the evolutionary tracks of galaxies in the $M_H-M_\star$ plane. Indeed, while for a quiescent galaxy there is no evolution of stellar mass ($M_{\star, i}(z)=M_{\star, i}(0)$), the host halo mass may evolve in time. Therefore its evolution in the $M_H-M_\star$ plane is influenced by the choice of the halo in which it is hosted.

\begin{figure*}
\centering
\includegraphics[scale=0.5]{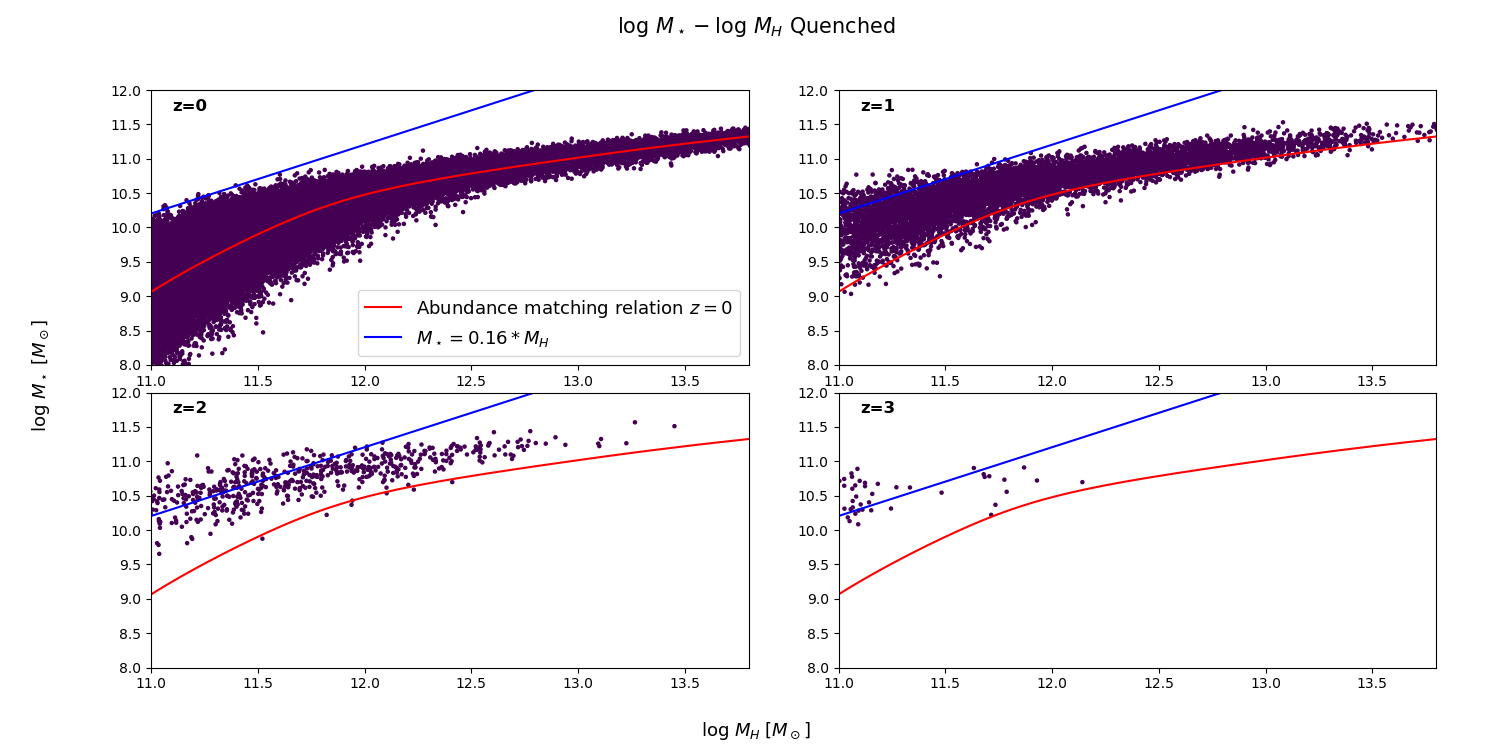}
\includegraphics[scale=0.5]{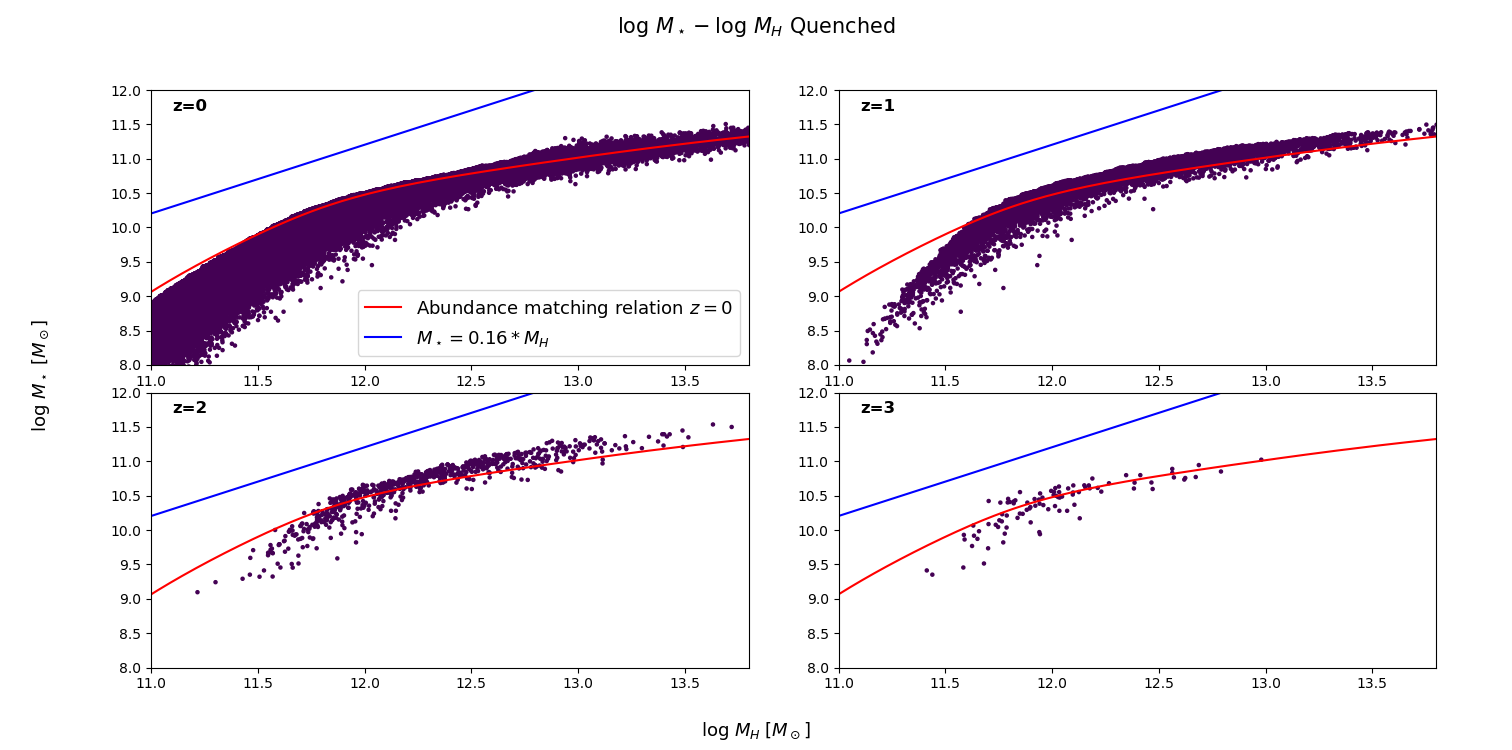}
\caption[]{SMHM relation for quenched galaxies at different redshifts: $z=0$ top left, $z=1$ top right, $z=2$ bottom left, $z=3$ bottom right. Violet points: galaxies from the catalogue, red line: SMHM relation from abundance matching at $z=0$, blue line: relation $M_\star=0.16\,M_H$. Top panel: quenched galaxies selected randomly, bottom panel: quenched galaxies selected by their distance from $M_\star=0.16\,M_H$ relation. We can notice that, at given redshift and stellar mass bin, a random selection of galaxies to activate between $z$ and $z+dz$ can give origin to some galaxies with above the blue line. We can avoid this issue by selecting galaxies to activate between the nearest to the $M_\star=0.16\,M_H$ relation.}
\label{fig:SMHM_quenched_distance}
\end{figure*}

The most agnostic method is to select quenched galaxies randomly. However, as we show in Figure \ref{fig:SMHM_quenched_distance}, it produces some unwanted features in the $M_H-M_\star$ plane. In the top panel of Figure \ref{fig:SMHM_quenched_distance} we show the evolution of quiescent objects in the $M_H-M_\star$ plane (top left $z=0$, top right $z=1$, bottom left $z=2$ and bottom right $z=3$). We can see that some quenched galaxies end up having a stellar mass $M_{\star,i}>M_{\star,i,max}(z)$ where $M_{\star,i,max}=0.16\,M_{H,i}(z)$ is the maximum stellar mass allowed by the average baryon to dark matter ratio (blue lines in the Figure). This behavior is due to our random selection criterion to activate galaxies going to higher redshift. Indeed, sometimes it happens that galaxies with stellar mass near the $M_{\star,i,max}(z)$ limit, remain quenched at higher redshift, with their stellar mass remaining constant and their halo mass which keeps decreasing, progressively reducing the mass limit $M_{\star,i,max}(z)$ to a value $<M_{\star,i}$. While this behavior is found also in some hydrodynamical simulations, we prefer to keep our mock galaxies below the $M_{\star,i,max}(z)$ limit and choose another selection criteria for quenched galaxies. 

In order to satisfy the condition $M_{\star,i}\leq M_{\star,i,max}(z)$, we should activate preferentially galaxies close to the mass limit. For this reason, at a given stellar mass, we activate the number of galaxies required by the factor $1-f_{Q-Q, \rm cat}$ selecting them among the nearest to $M_{\star,i,max}$. Since a quenched galaxy will have a horizontal evolution in the $M_H-M_\star$ plane, i.e., constant $M_\star$ and evolving $M_H$, this is the best way to ensure galaxies do not overshoot $M_{\star,i,max}$. In physical terms, we are requesting that haloes hosting quenched galaxies were already massive enough at the moment of quenching $z_Q$ to provide the necessary amount of gas to form a stellar mass $M_{\star,i}$. In the bottom panel of Figure \ref{fig:SMHM_quenched_distance} we show the evolution of quenched galaxies in the $M_H-M_\star$ plane for this selection criterion. We can see that they never overstep the $M_{\star,i,max}$ limit.

We close this section by stressing that the selection criterion for quenched galaxies is a degree of freedom for our model. We choose to preferentially quench galaxies based on their proximity to the $M_{\star,i,max}=0.16\,M_{H,i}(z)$ relation, but in principle any other method could be used, provided that it does not produce galaxies with $M_{\star,i}>M_{\star,i,max}(z)$. Many theoretical works have studied possible physical reasons for quenching relating it to DM/gas accretion, mergers and/or central SMBH activity. A possible future refinement of the model could be to implement theoretical prescriptions for galaxy quenching. Finally we note that, despite we assume as fiducial selection criterion the distance from $M_{\star,i,max}$, any other selection criterion would not produce substantial changes in the results we are going to show, apart for the evolution in the $M_H-M_\star$ plane. This is due to the fact that the sHAR distribution is independent of halo mass (see section \ref{sec:shar}). As a consequence, the backward evolution of a galaxy when it converts from quiescent to star-forming, is not dependent on the mass of the selected host DM halo. Therefore, the galaxy stellar mass assembly will be independent of the selection criterion, but the SMHM will change.

\newpage

\newpage

\subsection{Galaxy Stellar Mass Assembly}\label{sec:abundance_matching}
\begin{figure*}
\centering
\includegraphics[width=0.75\textwidth]{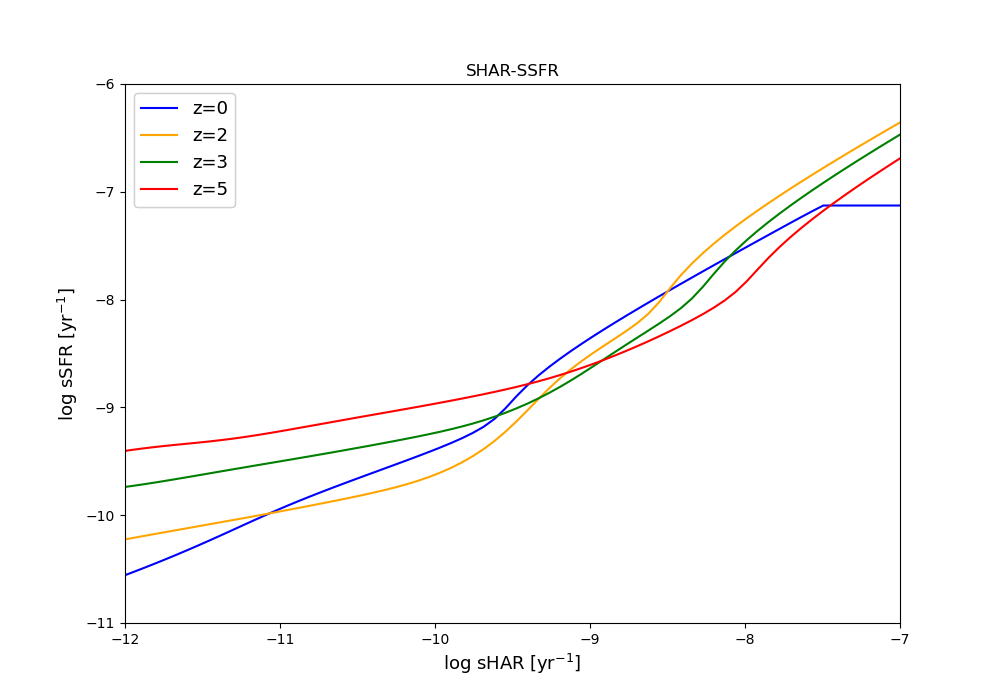}
\caption[]{sSFR-sHAR relation from abundance matching. Colors are as in Figure \ref{fig:HMF}.}
\label{fig:sSFR-SHAR relation}
\end{figure*}

Once we are able to empirically trace quenched galaxies, we need a way to follow the evolution of star-forming ones, which implies being able to assign a star formation history to them. In the empirical model under consideration we assume a monotonic relationship between sSFR and sHAR (equation \eqref{eq:abma_s}), which we derive via abundance matching without any a-priori parametrization. The statistical distributions used to perform abundance matching are (i) the probability distribution of sSFR, $\rm dp/dsSFR$, obtained normalizing equation \eqref{eq:dotNdVdsSFR}, and (ii) the probability distribution of sHAR for haloes containing star-forming galaxies. This means we should remove quenched galaxies from the computation of the sHAR functions in equation \eqref{eq:dpdlogSHAR}. This can be easily done as:

\begin{equation}
\begin{split}
\frac{\rm d^2N_{\rm ac}}{\rm dV\,d\log sHAR}=&\frac{1}{\rm N_{\rm halo}}\,\frac{\rm dN}{\rm dV}\times\\
&\times\sum_j{\delta(\rm \log sHAR-\log sHAR_{j})\,p_j},
\end{split}
\label{eq:dpdlogSHAR_ac}
\end{equation}
where $j$ runs only over haloes containing galaxies selected as star-forming. The sHAR probability distribution for haloes hosting active galaxies ($\rm dp_{\rm ac}/d\log sHAR$) is just the normalization of equation \eqref{eq:dpdlogSHAR_ac}. Abundance matching can be performed by solving the equation:

\begin{equation}
\begin{split}
\int_{\log\rm sSFR}^{\infty} &\,\rm d\log sSFR'\,\frac{\rm dp}{\rm d\log sSFR'}=\\
&=\int_{-\infty}^{\infty}\,d\log sHAR'\,\frac{\rm dp_{\rm ac}}{\rm d\log sHAR'}\times\\
&\times\frac{1}{2}\,\rm erfc \left[\frac{\log(\rm sHAR(sSFR)/sHAR')}{\sqrt{2}\,\sigma_{\log\rm sSFR}}\right],
\end{split}
\label{eq:abma_s_eq}
\end{equation}
where $\sigma_{\log\rm sSFR}$ is the scatter on the relation. The resulting $\rm sSFR=sSFR(z,sHAR)$ relation is shown in Figure \ref{fig:sSFR-SHAR relation}, where different colors stand for different redshifts. By construction, it is monotonic with a normalization tending to increase with redshift at lower sHAR and to decrease at higher sHAR. The shape of the relation is similar to a double power law, where the steeper part reflects the presence of the starburst population.

As stated in section \ref{sec:initialization}, we initialize our galaxies on the $z=0$ SMHM relation and we evolve them backwards in time. We select quenched galaxies as shown in section \ref{sec:quenching} and we assign a sSFR only to star-forming ones. The sSFR of the $i-$th galaxy of the catalogue will be: $\rm sSFR_i(z)=sSFR(z,sHAR_i)$$+r$, where $r$ is just a random number drawn from a log-normal distribution with scatter $\sigma_{\log\rm sSFR}$ to keep into account the scatter around the relation. With this procedure we obtain a sSFR for each galaxy in the catalogue at all redshifts. The evolution of stellar mass and SFR of each galaxy can be obtained by:

\begin{align}
    &\rm M_\star(\rm z)=M_0\,\exp{\left[-(1-\mathcal{R})\,\int_{z}^{0}\,\rm dz'\,\frac{\rm dt}{\rm dz'}\,sSFR(z')\right]}\label{eq:mass_growth}\\
    &\psi(z)=\frac{1}{1-\mathcal{R}}\,\frac{\rm dM_\star}{\rm dt},\label{eq:psi}
\end{align}
where $M_0$ is the stellar mass at $z=0$ and $\mathcal{R}\simeq 0.44$ is the recycling gas fraction.

\begin{figure*}
\centering
\includegraphics[scale=0.55]{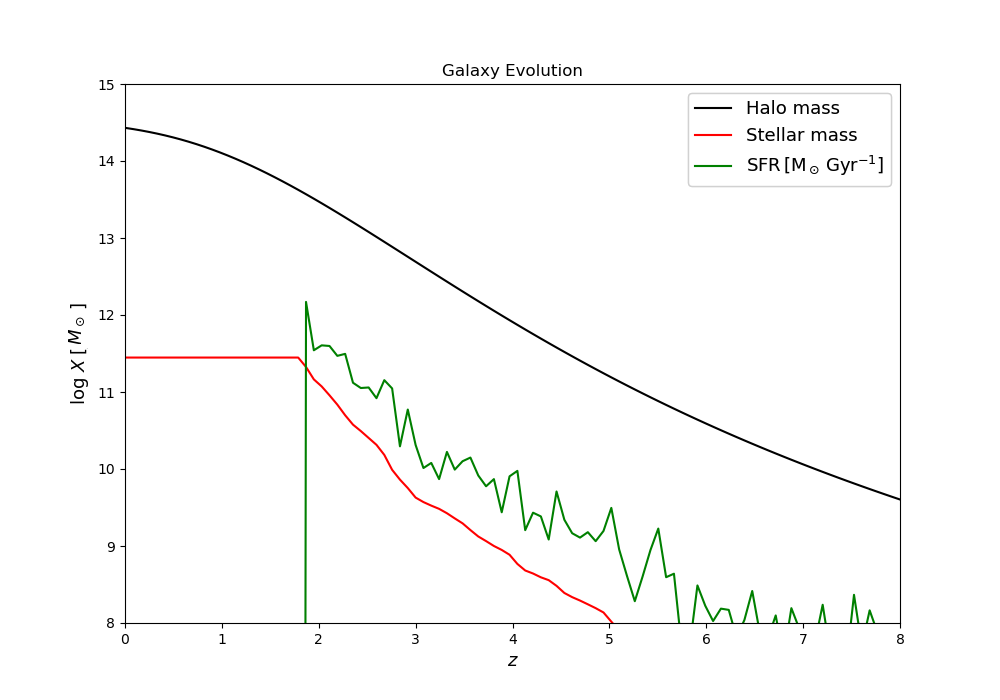}
\includegraphics[scale=0.55]{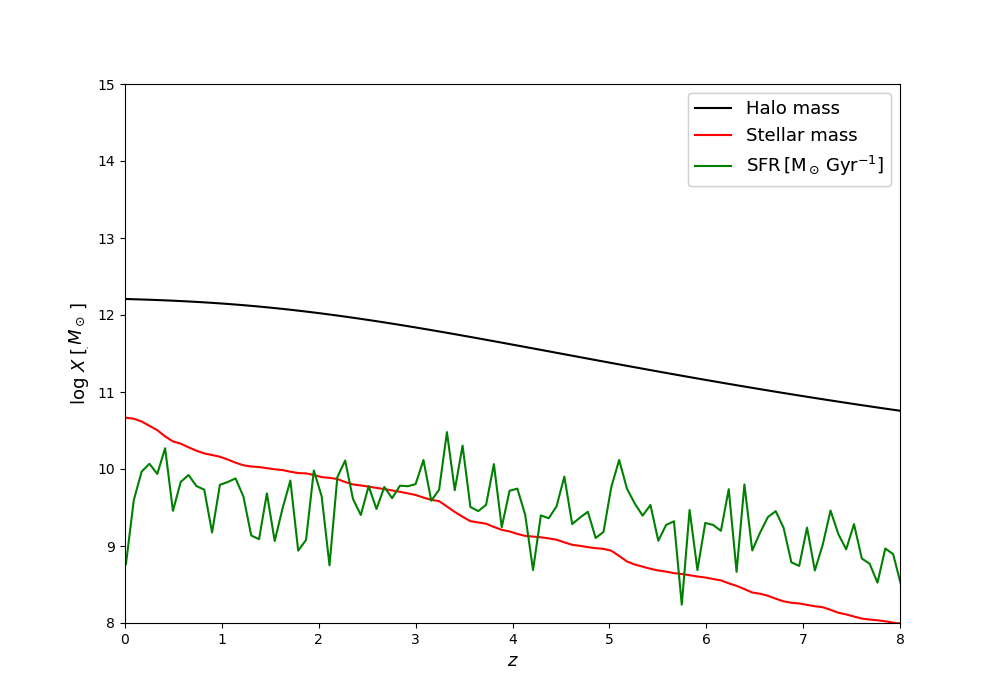}
\caption[]{Example of evolution of 2 mock galaxies. Black line: halo mass, red line: stellar mass, green line: SFR in $\rm M_\odot/Gyr$. Top panel shows a quenched galaxy which formed all its stellar mass at $z>2$ with large SFR, bottom panel a star-forming one with low and prolonged SFR.}
\label{fig:evolution}
\end{figure*}

Examples of the evolution of some galaxies in our catalogue are shown in Figure \ref{fig:evolution}, where we can see the halo mass and stellar mass growth, as well as the SFR as a function of redshift for 2 different kinds of galaxies: a massive quenched galaxy residing in a massive halo featuring a high level of star formation $\psi\geq 100\,\rm M_\odot\,yr^{-1}$ up to the redshift of quenching $z_Q\sim 2$, and a star-forming one residing in a smaller halo and featuring a lower level of star formation $\psi\sim 5\,\rm M_\odot\,yr^{-1}$ but forming stars for all its lifetime.

\section{Model Results}\label{sec:results}
Our basic and transparent model presented above is fully driven by empirically determined relations. It is built only on two main assumptions: a selection criterion for quenched galaxies and the monotonicity between sSFR and sHAR for star-forming ones. The actual number of parameters is just two: the scatter around the SMHM at $z=0$, which is our initial condition, and the scatter around the sSFR-sHAR relation. All the other quantities appearing in the equations are directly taken from observational data and do not imply any assumption or theoretical modelization. Despite its simplicity, the model is able to reproduce well the evolution of the stellar population in galaxies, at least in a statistical sense, as we will show in the next sections.

\subsection{Stellar Mass Functions}
\begin{figure*}
\centering
\includegraphics[scale=0.7]{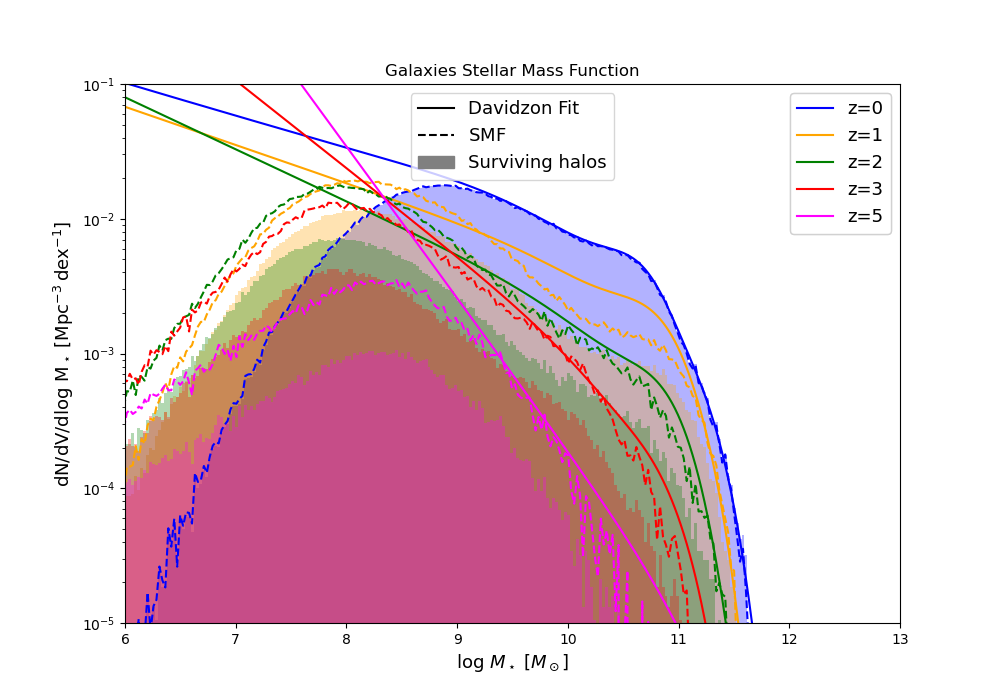}
\caption[]{Total stellar mass function evolution with redshift: blue $z=0$, orange $z=1$, green $z=2$, red $z=3$ and magenta $z=5$. Histograms are the SMF for surviving haloes, dashed lines the total SMF reconstructed from our catalogue and solid lines the fit by Davidzon et al. (2017). Our model is able to well reproduce the evolution of SMF up to $z\sim 5$.}
\label{fig:SMF_model}
\end{figure*}

In this section we show the goodness of the model in reproducing the statistical evolution of galaxies, traced by the stellar mass functions. While the stellar mass functions for quiescent objects are reproduced by construction, as shown in Figure \ref{fig:SMF_quenched_distance}, the total stellar mass function at different redshifts is a genuine prediction of our model, since it is based on the evolution of our mock galaxies and thus on the derived sSFR-sHAR relation. In Figure \ref{fig:SMF_model} we show, at different $z$ (color code), the comparison between the mass function for surviving haloes in our catalogue (histograms), the predicted SMF for all the galaxies (dashed lines), derived by weighting each halo with the factor $p_i$, and the Davidzon et al. (2017) fit (solid lines). Since at $z=0$ our initial condition assigns stellar masses on the basis of the SMHM, the SMF at $z=0$ is reproduced by construction. However, our model is able to fairly trace the SMF evolution at all redshifts up to $z\sim 5$, where we start having a lack of statistics at high stellar masses. The only significant deviation between the model and the fit to observational data is at $z\sim 1$ where the model seems to predict an excess of galaxies with low mass $\log M_\star\lesssim 9.5$ and a deficit of galaxies with $\log M_\star\geq 9.5$, even though the maximum difference is always smaller than a factor $1.5$. The good match with the global SMF at all redshifts and stellar masses further supports our assumption of monotonicity between sSFR and sHAR.

\subsection{Evolution of SMHM}
\begin{figure*}
\centering
\includegraphics[scale=0.52]{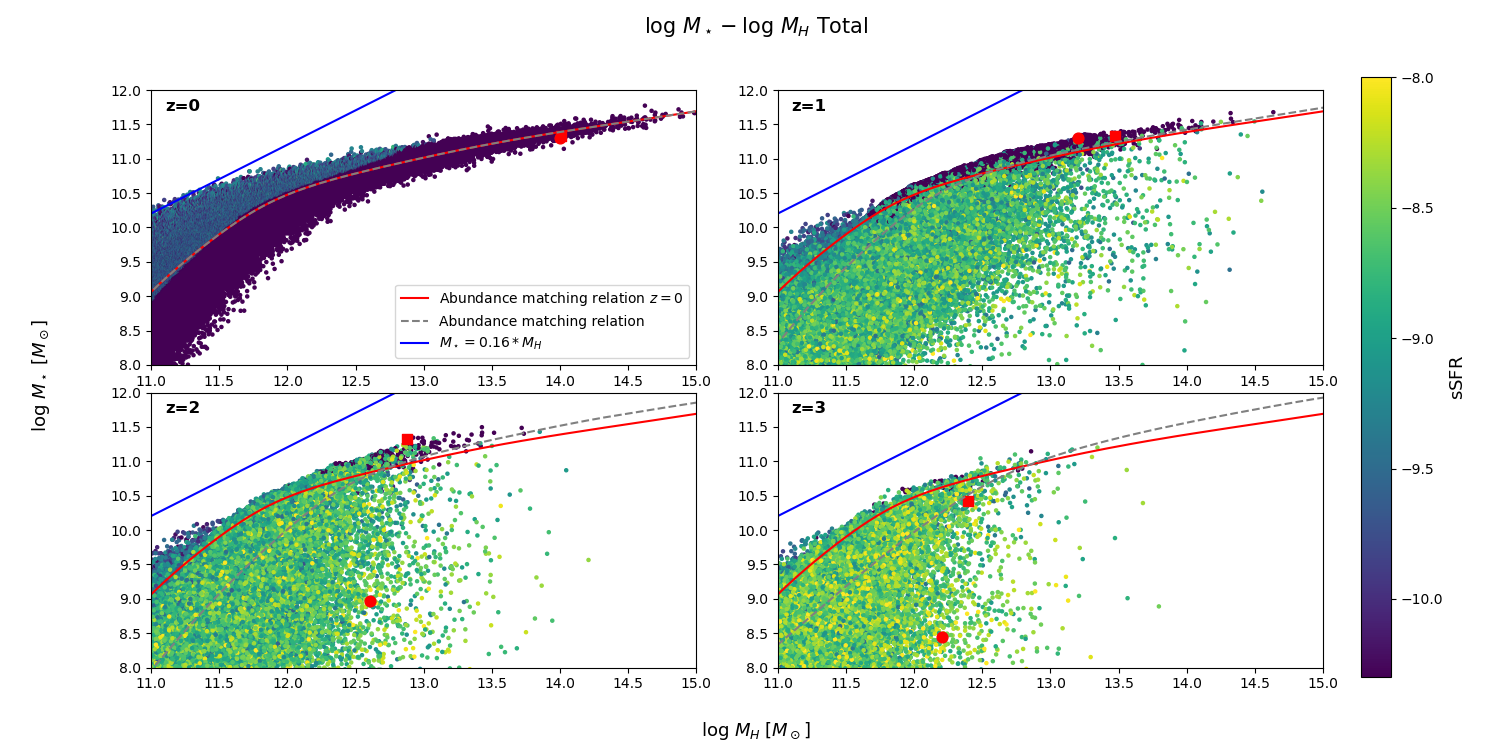}
\caption[]{SMHM relation for all galaxies in the catalogue at different redshifts: $z=0$ top left, $z=1$ top right, $z=2$ bottom left, $z=3$ bottom right. The colour of the dot represents the sSFR. The big red dot is the evolution of 1 galaxy as an example. The solid red line represent the abundance matching relation at $z=0$, while the grey dashed line the derived average $M_H-M_\star$ relation at the specific redshift of the panel. The SMHM is an output of our model.}
\label{fig:SMHM_model}
\end{figure*}

A natural prediction of the empirical model is the evolution of the relation between $M_\star$ and $M_H$ at all redshifts. Given the $z=0$ SMHM with a scatter, which is our initial condition, we can predict its evolution backwards in time. The result is shown in Figure \ref{fig:SMHM_model}. Points represent the positions of our mock galaxies at different redshifts ($z=0$ top left, $z=1$ top right, $z=2$ bottom left, $z=3$ bottom right) and the color code represents their sSFRs (purple points are quenched galaxies). sSFRs are typically higher at high redshift due to the larger sHAR of DM haloes. The fraction of quenched galaxies increases going towards lower redshifts. Star forming galaxies evolve towards the top right part of the plot until they are quenched and start a horizontal evolution at fixed stellar mass and increasing halo mass. The average relation between $M_H$ and $M_\star$, obtained from abundance matching between stellar and halo mass, is shown as a grey dashed line. We can see that high-z galaxies tend to follow this relation, but with a quite large dispersion. Specifically, galaxies are not distributed symmetrically around it, but they tend to feature a long tail toward low stellar masses at fixed halo mass. The $1-\sigma$ scatter at $z=3$ can be as high as $\sim 1\,\rm dex$. We expect that this result is partly driven by our assumptions and partly by physical reasons. On the one hand, the initial condition imposed at $z=0$ influences the scatter even at high-z: a less/more scattered $z=0$ relation would produce a less/more scattered high-z distribution. On the other hand, scatter is naturally originated by the fact that we are not matching stellar and halo masses, but sHAR and sSFR, so predicting the evolution of galaxies in the $M_\star-M_H$ plane. The scatter is due to the different assembly time of galaxies. This can be seen from the big red circle and square, representing the evolutionary tracks of two example galaxies. While the two galaxies have the same stellar and halo mass at $z=0$, they had different quenching times (the circle quenches at $1<z<2$ and the square at $2<z<3$) and a different evolution before quenching, with the circle having a fast growth at $z<2$ and the square forming most of its mass at $z>3$. 

\section{Bulge and Disk Formation}\label{sec:bulge_disk}
In the previous sections we have described our model which connects halo and galaxy properties and we have shown its ability in predicting some properties of evolving galaxies such as the build up of their stellar mass. Now we want to exploit the model to test a hypothesis for the morphological evolution of galaxies. The idea is extremely simple: we assume that the galactic bulge and spheroids are formed during the phase of fast accretion of dark matter, while disks are originated in the subsequent slow accretion phase. 

The idea is inspired by the fact that, as stated in sections \ref{sec:introduction} and \ref{sec:haloes}, during fast accretion DM haloes are growing the bulk of their mass through major mergers or strong inflows. These violent accretion processes may lead to a quick loss of angular momentum also for baryonic matter, e.g. through dynamical friction between giant gas clumps (see Lapi et al. 2018a; Pantoni et al. 2019) that can fastly sink to the very central region and originate the bulge. During slow accretion, instead, DM is mostly accreted via minor mergers, weak and steady DM flows or pseudo-evolution which typically contribute in building up the outskirts of the halo. For these reasons we believe baryons accreted during this phase are not directly funneled toward the center, but they mantain their angular momentum and start originating a more extended stellar disk. Such a hypothesis can qualitatively explain the differences in the age of the stellar population between (classical) bulge/spheroids and (thin) disk, with bulge stars being typically older, $\alpha-$enhanced and almost coheval and disk stars formed over a longer timescale. 

\begin{figure*}
\centering
\includegraphics[scale=0.55]{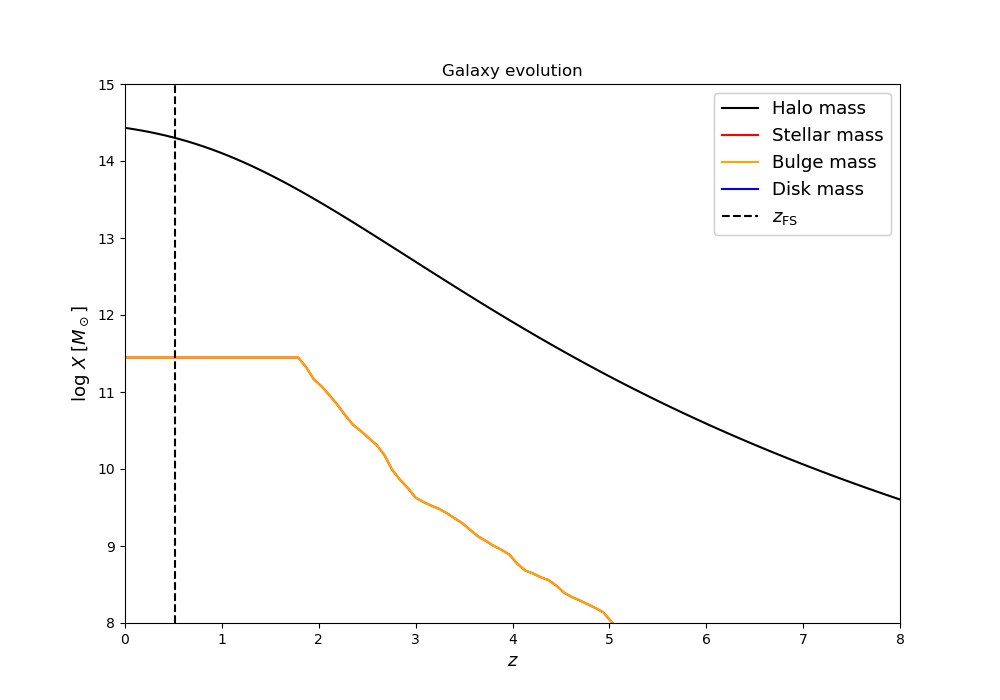}
\includegraphics[scale=0.55]{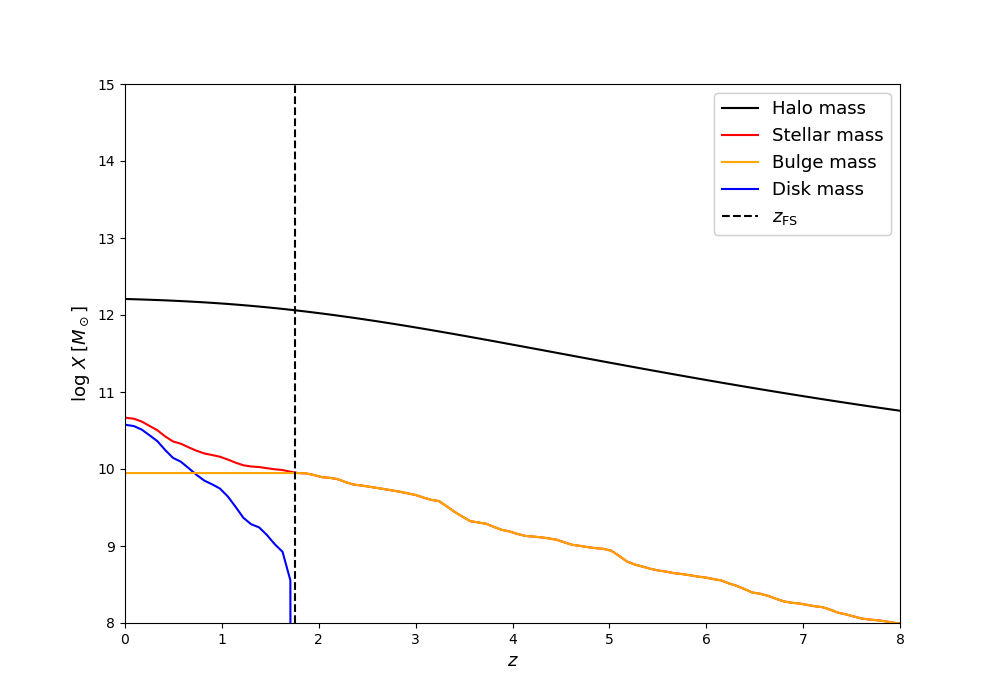}
\caption[]{Example of evolution of bulge and disk for 2 galaxies. Black line: halo mass, red line: total stellar mass, orange line: bulge mass, blue line: disk mass. Dashed vertical line represents the moment of transition between fast and slow accretion $z_{\rm FS}$. Top panel shows a quenched galaxy with $z_Q>z_{\rm FS}$, all the stellar mass in the bulge and the galaxy is an elliptical. Bottom panel shows a star-forming galaxy building its bulge up to $z_{\rm FS}\simeq 1.5$ and the disk at lower redshift.}
\label{fig:evolution_bd}
\end{figure*}

This idea is not completely new: it was exploited by Cook et al. (2009) who built a semi analytical model based on this assumption, finding encouraging results. We now want to test this hypothesis by the use of our simple and almost parameter-free semi empirical model. Given a transition redshift between fast and slow accretion $z_{\rm FS}$ for each host halo (defined in section \ref{sec:haloes}), we are able to predict the amount of stellar mass in the bulge $M_{\star,b}$ and in the disk $M_{\star,d}$ for each galaxy in the catalogue:

\begin{align}
    &M_{\star,b}(z)=(1-\mathcal{R})\,\int_{\infty}^{\max{(z,z_{\rm FS})}}\,\rm dz'\,\frac{\rm dt}{\rm dz'}\,\psi(z')\label{eq:mass_bulge}\\
    &M_{\star,d}(z)=(1-\mathcal{R})\,\int_{\max{(z,z_{\rm FS})}}^{z}\,\rm dz'\,\frac{\rm dt}{\rm dz'}\,\psi(z').\label{eq:mass_disk}
\end{align}
There are three possible scenarios that can occur, depending on the relative position of the transition redshift $z_{\rm FS}$ and the redshift at which quenching occurs $z_Q$:
\begin{itemize}
\item $z_{\rm FS}<0$. There are some cases in which haloes, especially massive ones, are still in the fast accretion regime at $z=0$. In these cases all the stellar mass formed is in the bulge, i.e., the galaxy is an elliptical.
\item $z_{\rm FS}>0$ and $z_Q>z_{\rm FS}$. In this case the transition between fast and slow accretion occurs before $z=0$, but the galaxy has already quenched before the transition. Also in this case all the stellar mass formed is in the bulge and the galaxy is an elliptical.
\item $z_{\rm FS}>0$ and $z_Q<z_{\rm FS}$. In this case the transition between fast and slow accretion occurs before $z=0$ and quenching happens after the transition or does not happen at all. In this case all the stellar mass formed before the transition is in the bulge and the rest in the disk. The relative abundance of bulge and disk components are determined by the precise values of $z_Q$ and $z_{\rm FS}$: an early transition and no quenching would result in a galaxy with a prominent disk; a late transition, close to the quenching time, would result in a lenticular galaxy.
\end{itemize}
Two examples of the SFH of galaxies as a function of redshift can be seen in Figure \ref{fig:evolution_bd}. The bulge mass growth is shown in orange, the disk mass in blue and the total stellar mass in red. In the top panel we show a case in which the fast-slow transition occurs after quenching; therefore all the stellar mass is in the bulge. In the bottom panel we show the case of a galaxy with no quenching; the separation between the bulge and disk components is evident.

One of the most striking differences between different morphological types of galaxies is the relation between morphology, stellar mass and star formation activity. Elliptical galaxies, in the local Universe, tend to be more massive and their star formation activity is so low they are classified as quiescent. Disk galaxies, instead, are generally less massive than ellipticals and they are star-forming at the present time. Such a strong bimodality is the main signature of the existence of a link between the quenching process and galaxy structure evolution (Wuyts et al., 2011; Whitaker et al., 2015; Dimauro et al., 2022). In our model, quenching is not directly related to morphology, but a link between them naturally emerges. Indeed, the transition redshift $z_{\rm FS}$, on average, decreases for higher descendant halo masses. Consequently, massive haloes, typically hosting massive quiescent galaxies, have less or no time to develop a substantial stellar disk and they tend to host pure spheroidals. Contrariwise, star-forming galaxies are usually hosted in lower mass haloes, which have a higher transition redshift and more time to develop a stellar disk.

\begin{figure*}
\centering
\includegraphics[width=0.75\textwidth]{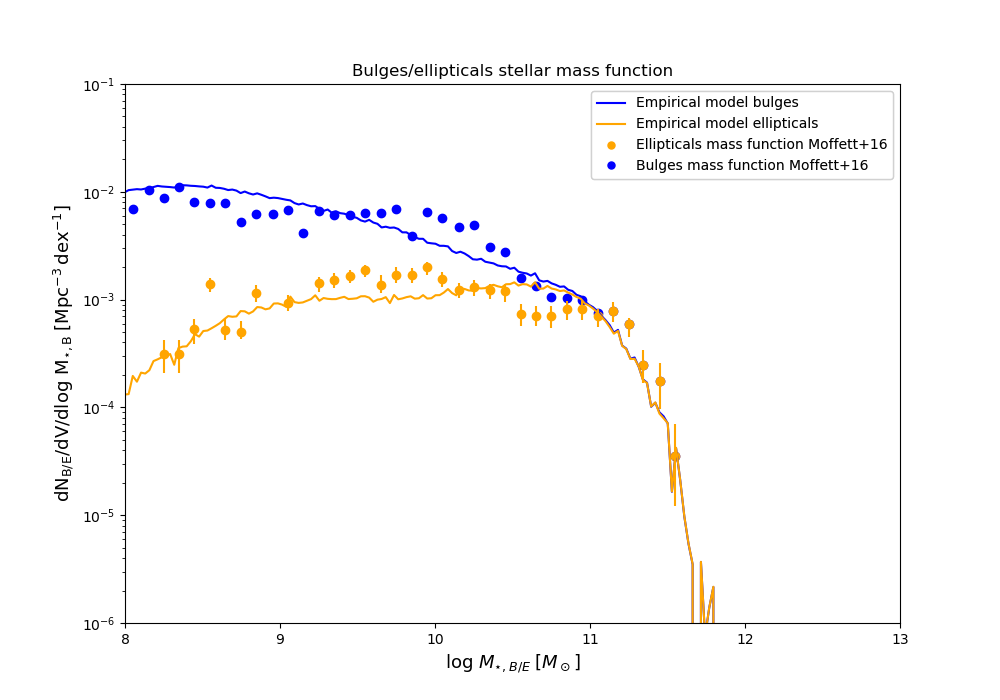}
\caption[]{Bulge stellar mass function. Solid lines are model predictions, dots are GAMA data. In blue the total bulge stellar mass function, in orange the stellar mass function for ellipticals.}
\label{fig:SMF_B}
\end{figure*}

It is important now to statistically evaluate the hypothesis for disk and bulge formation comparing it with the available data for galaxy morphology. We construct the stellar mass function only for bulges at $z=0$ and compare it with the observational GAMA data from Moffett et al. (2016b). This comparison is shown in Figure \ref{fig:SMF_B} where the blue and orange lines are our predicted stellar mass functions for all the bulges and for ellipticals, defined as galaxies with bulge over total ratio $\rm B/T>0.7$ , a classical choice often adopted in semi analytic models and observational works (Weinzirl et al. 2009; Wilman et al. 2013; Fontanot et al. 2015; Henriques et al. 2020). The agreement with GAMA data is impressive for both ellipticals and bulges. 

\begin{figure*}
\centering
\includegraphics[width=0.75\textwidth]{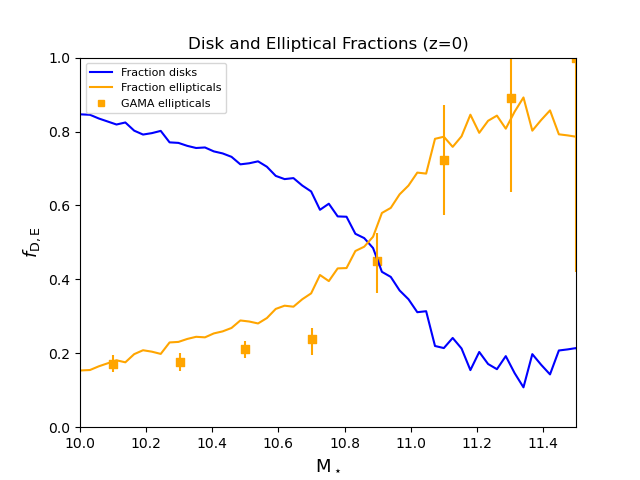}
\caption[]{Fraction of disks and ellipticals as a function of stellar mass at $z=0$. Solid lines are the model results, square dots GAMA data. Orange stands for the ellipticals fraction, while the blue line is the disk fraction and its just the reciprocus of the orange line.}
\label{fig:Efraction_M}
\end{figure*}

In Figure \ref{fig:Efraction_M} we show instead the fraction of elliptical galaxies as a function of the stellar mass at $z=0$ (orange line). We can see that such a fraction increases with $M_\star$, ranging from $\sim 0.15$ at $M_\star\sim 10^{10}\,M_\odot$ to $\sim 0.75$ at $M_\star\sim 10^{11}\,M_\odot$. Even in this case the agreement with GAMA data is very good, at least up to $M_\star\gtrsim 10^{11}\,M_\odot$. At higher stellar masses the fraction of ellipticals remains high but the trend becomes unclear. This is due at least to two reasons: (i) lack of statistics, i.e., at high stellar masses the number of galaxies per mass bin becomes very low (of the order of unity) and this creates an enormous scatter on the results; (ii) dry mergers become important and may contribute in shaping the morphology of galaxies. 

These results indicate that our hypothesis linking the bulge/disk dicothomy to different halo accretion modes is very promising, with an early formation of the bulge or spheroidal component and a subsequent generation of a stellar disk. In our model, disk galaxies start to dominate at $z\lesssim 1$, since the average transition redshift between fast and slow accretion $z_{\rm FS}$ ranges between $z_{\rm FS}\sim 1$ for haloes with small descendant mass $M_H\sim 10^{11}$ and $z_{\rm FS}\sim 0$ for haloes with large descendant mass $M_H\sim 10^{15}$, even though the large scatter in $z_{\rm FS}$ allows for the early formation of some disk galaxies. Such a morphological evolution is in agreement with stellar population studies of early type galaxies (Thomas et al. 2005, 2010; Gallazzi et al. 2006; Bellstedt et al. 2023) and late type ones (Pezzulli \& Fraternali 2016; Grisoni et al. 2017). However, recent JWST observations with NIRCam (Ferreira et al. 2022a, 2022b) show the early formation of a disk in high redshift galaxies, with a significant fraction of disk-like morphologies $\sim 40\%$ out to $z\sim 4$, which may seem at variance with the standard paradigm. Nevertheless, these high redshift disks could substantially differ from low redshift ones, being generally less extended in size, thick, compact, clumpy and possibly subject to instabilities and compaction events, which may trigger a central starburst leading to the bulge formation, as suggested by several theoretical models (e.g. Barro et al. 2013; Dekel \& Burkert 2014; Lapi et al. 2018), simulations (e.g. Johansson et al. 2012b; Zolotov et al. 2015; Lapiner et al. 2023) and supported by observations (e.g. van Dokkum et al. 2015; Wisnioski et al. 2018; Talia et al. 2018). A possible explanation is that current JWST NIRCam observations probe the rest frame optical emission of high redshift galaxies, but may suffer from strong dust attenuation, especially in the central regions (Cheng et al. 2022; Shen et al. 2023). Therefore, these results may probe fairly well the optical emission from the stellar disk, but miss the compact far-IR emission from dust in the central regions, possibly associated to the central starburst, which could indicate a bulge caught in the act of formation. The necessity of multi-band observations to establish galaxy properties and morphologies has clearly emerged already in the pre-JWST era with the advent of ALMA, which revealed the presence of compact $\lesssim 1\,\rm kpc$ dust-enshrouded cores with extremely high SFR $\psi\sim100-1000\,\rm M_\odot\,yr^{-1}$ in the central regions of high redshift galaxies that were completely missed in optical/UV observations (Tadaki et al. 2017a, 2017b, Fujimoto et el. 2017; Gullberg et al. 2019). These authors have shown that these central cores can be up to $\gtrsim 10$ times smaller and yield a SFR $\sim 10-100$ times larger than the extended optical/UV emission from the outer disk observed with HST. For these reasons, some of the high-z galaxies visually classified as disks or peculiar could actually be bulges in the act of formation, with some optical emission coming from a less obscured extended disk with moderate star formation. Therefore, a firm conclusion on these data may be drawn only after they are complemented with multiwavelength observations, especially in the far-IR/submm band.

A related point concerns disk disruption and regrowth: hydrodynamical simulations show that galaxies tend to destroy their disk via disruptive events as major mergers and, in some instances, regrow it subsequently. This could imply that the population of high redshift disks is not representative of the progenitors of local stable disks. In our two-phase model, we do not follow in fine details the regrowth of the galactic disk between major mergers, but we simply assume that all the stars formed during fast accretion end up being locked up in a bulge. The rationale under this assumption is that major mergers, instabilities and disruptive events are frequent during fast accretion and, while it is still possible that some disks temporarily form, they are rapidly washed out by such disruptive events. Disk regrowth is instead present during slow accretion, since this is dominated by minor mergers and smooth accretion, so the disk has time to grow and stabilize around the early formed bulge. Depending on the transition redshift and on the redshift of quenching, disk regrowth can be more or less important.

All in all, we have demonstrated that our semi empirical model, albeit treating some processes in a simplified way, is able to trace galaxy evolution with a very small set of assumptions and can be extremely useful in testing physical hypotheses, since their effect can be directly linked to observables without degeneracies with other parameters.

\section{Conclusion}\label{sec:conclusion}
The complex topic of galaxy formation and evolution is still a matter of intense investigation, as many aspects are still controversial and largely unsolved. In this paper we have presented a new semi empirical model to study galaxy evolution, expressively designed on very few assumptions and free parameters. The main novelty points are three:
\begin{itemize}
\item We treat quenched galaxies in a purely empirical way, not trying to explain the quenching mechanism but just tracing their number and their evolution directly from the SMF. In this way we can test our hypothesis on the SFR evolution in a simple and transparent way, marginalizing out quiescent objects.
\item We do not make use of the standard $M_H-M_\star$ abundance matching; rather we match the specific halo accretion rate with the specific star formation rate of the galaxy. The reason for this choice is that masses are integrated quantities, depending on the overall accretion history, which might introduce some spurious dependence in the $M_H-M_\star$ relation, such as assembly bias. Instead sHAR and sSFR are more related to the situation at a specific time. This assumption would lead to a prediction for the evolution of the SMHM relation and its scatter.
\item We use the model to test a hypothesis on the morphological evolution of galaxies: we suggest that the bulge/disk dicothomy observed in galaxies is originated by 2 different phases in the DM halo accretion (fast and slow accretion). During fast accretion, dominated by major mergers and/or violent accretion the halo potential well is continuously reshuffled; we prescribe that the bulge/spheroidal component is formed in this phase. The subsequent slow accretion phase is instead dominated by steady accretion on the halo outskirt, not affecting the central potential well; we believe in this phase is formed a more extended rotating stellar disk.
\end{itemize}

The model is extremely simple and based on solely 2 assumptions and 1 initial condition, with only 2 free parameters. The main assumption is the monotonic relation between sSFR and sHAR which is parametrized by 1 free parameter $\sigma_{\log\rm sSFR}$, the scatter on the relation (for more details see section \ref{sec:empirical_model}). The second assumption is a selection criterion for quenched galaxies , which we have proven not to alter the main results of the model (see section \ref{subsec:selection_criterion} for more details). Finally, since the model associates a sSFR to each halo, we need to know the stellar mass at some moment in time to obtain a star formation history. We impose this initial condition at $z=0$ where our data are more robust and complete. We assign stellar masses to haloes by the use of the SMHM at $z=0$ and then we evolve our galaxies backwards in time. This relation is parametrized by the second parameter $\sigma_{M_\star}$, which is the second and last parameter.

We prove in section \ref{sec:results} that this simple model is able to excellently reproduce the evolution of the galaxies SMF up to $z\sim 5$. This is an important result; indeed, given the very few parameters of the model, the reproduction of the SMF directly tests the goodness of the assumed monotonicity between $\rm sSFR$ and $\rm sHAR$. We also predict the evolution of the SMHM relation at $z>0$, finding that such a relation is rather disperse, possibly due to the fact that masses are integrated quantities and their value depends on many other parameters related to the accretion history such as the time of halo formation, the time at which substantial star formation activity starts and the time of quenching.     

We then refine the model to keep into account our hypothesis for morphological evolution following these 2 steps:
\begin{itemize}
\item We divide the DM halo growth in a phase of fast accretion at early times and slow accretion at late times (see section \ref{sec:haloes}).
\item We assign all the stellar mass formed during fast accretion to the bulge/spheroidal component and all the stellar mass formed at late times to the disk (section \ref{sec:bulge_disk}).
\end{itemize}
Again, the flexibility of a semi empirical model with very few parameters comes to good use since we can directly test this physical hypothesis without tuning any parameter. We find an excellent agreement between our predicted bulge stellar mass function and the observational determination by Moffett et al. (2016). We are also able to well reproduce the stellar mass function for elliptical galaxies and their fraction as a function of stellar mass. 

All in all, the extreme semplicity of the semi empirical model allowed us to test 2 physical hypotheses in a transparent way: the sSFR-sHAR correlation and the bulge/disk-fast/slow accretion dicothomy. On top of this backbone, possible future refinements/extensions are in order and can go in many directions: (i) implementations of some physical recipe for quenching, which on the one hand should be in agreement with the empirical SMF, on the other hand it can naturally provide a selection criterion to quench galaxies; (ii) implementation of a mechanism for the growth of the central SMBHs (e.g. exploiting the mechanism presented in Boco et al. 2020, 2021a), relating it to some observational quantity; (iii) modeling the behavior of cold and hot gas; (iv) study the metal enrichment history in bulges and disks; (v) exploiting the mock catalogue of galaxies to predict compact binary merger rates to refine and extend the work of Boco et al. (2019, 2021b) 

With the advent of JWST, which will refine and extend our knowledge of the stellar mass function and luminosity function for galaxies, we believe that the future for semi empirical models is shining bright.

This work has been supported by the EU H2020-MSCA-ITN-2019 Project 860744 ‘BiD4BESt: Big Data applications for black hole Evolution STudies’. We thank the anonymous referee for constructive comments, which helped improving the final version of the manuscript. LB thanks Luigi Bassini, Tommaso Ronconi, Massimiliano Parente and Andrew Hearin for helpful discussions. AL acknowledge funding from the PRIN MIUR 2017 prot. 20173ML3WW, ‘Opening the ALMA window on the cosmic evolution of gas, stars and supermassive black holes’.

\end{document}